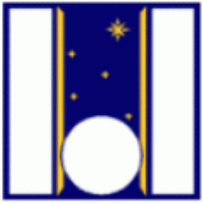

Fundación Galileo Galilei - INAF
Telescopio Nazionale Galileo

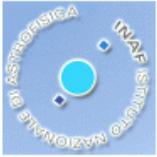

Istituto Nazionale di Astrofisica - Roma

# TNG publications
# 1989 – 2005

by **Walter Boschin**

Fundación Galileo Galilei - INAF

Last update: April 12, 2006

Santa Cruz de La Palma – October 2005

# Introduction

This document lists a set of (refereed and unrefereed) scientific publications based on data taken with the instruments of the Italian **Telescopio Nazionale Galileo** (**TNG**, mainly from the year 2000 onward) and the technical papers describing the development of the TNG project from the "phase A" (late '80s) until the end of year 2005. The collection is compiled by searching for publications on the internet. In particular, the search engines of the *NASA Astrophysics Data System* and *Google Scholar* are used.
This work represents the first attempt to probe the scientific production of the TNG and will be updated regularly from year to year. Comments and suggestions are welcome.

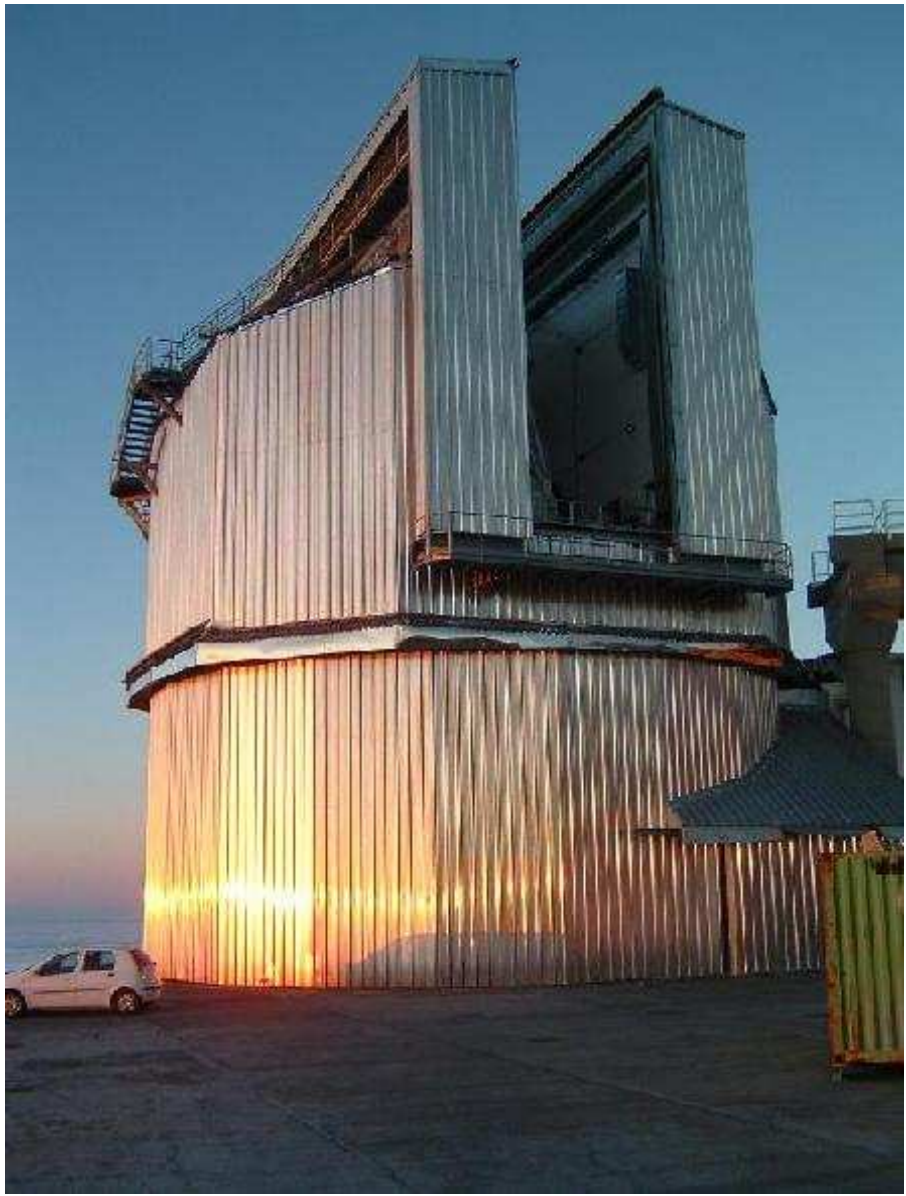

# Publications 2005 on international journals with "referee"

**Aguerri, J. A. L.; Elias de la Rosa, N.; Corsini, E. M.; Muñoz Tunon, C.**, "Photometric properties and origin of bulges in SB0 galaxies", 2005, A&A, 434, 109 (partly based on observations made with DOLoRes@TNG)

**Barrena, R.; Ramella, M.; Boschin, W.; Nonino, M.; Biviano, A.; Mediavilla, E.**, "VGCF detections of galaxy systems at 0.2<z<0.6 with optical multi-band observations", 2005, A&A, 444, 685 (partly based on observations made with DOLoRes@TNG)

**Barucci, M. A.; Fulchignoni, M.; Fornasier, S.; Dotto, E.; Vernazza, P.; Birlan, M.; Binzel, R. P.; Carvano, J.; Merlin, F.; Barbieri, C.; Belskaya, I.**, "Asteroid target selection for the new Rosetta mission baseline. 21 Lutetia and 2867 Steins", 2005, A&A, 430, 313 (partly based on observations made with DOLoRes and NICS@TNG)

**Battinelli, P.; Demers, S.**, "Exploratory C star search in the disk of M 31 beyond 30 kpc", 2005, A&A, 430, 905 (based on observations made with DOLoRes@TNG)

**Beirao, P.; Santos, N. C.; Israelian, G.; Mayor, M.**, "Abundances of Na, Mg and Al in stars with giant planets", 2005, A&A, 438, 251 (partly based on observations made with SARG@TNG)

**Bellazzini, M.; Gennari, N.; Ferraro, F. R.**, "The red giant branch tip and bump of the Leo II dwarf spheroidal galaxy", 2005, MNRAS, 360, 185 (based on observations made with DOLoRes@TNG)

**Bragaglia, A.; Held, E. V.; Tosi, M.**, "Radial velocities and membership of stars in the old, distant open cluster Berkeley 29", 2005, A&A, 429, 881 (based on observations made with DOLoRes@TNG)

**Capria, M. T.; Cremonese, G.; Bhardwaj, A.; de Sanctis, M. C.**, "O(S) and O(D) emission lines in the spectrum of 153P/2002 C1 (Ikeya-Zhang)", 2005, A&A, 442, 1121 (based on observations made with SARG@TNG)

**Cesaroni, R.; Neri, R.; Olmi, L.; Testi, L.; Walmsley, C. M.; Hofner, P.**, "A study of the Keplerian accretion disk and precessing outflow in the massive protostar IRAS 20126+4104", 2005, A&A, 434, 1039 (partly based on observations made with NICS@TNG)

**Chiaberge, M.; Sparks, W. B.; Macchetto, F. D.; Perlman, E.; Capetti, A.; Balmaverde, B.; Floyd, D.; O'Dea, C.; Axon, D. J.**, "The Infrared-dominated Jet of 3C 401", 2005, ApJ, 629, 100 (partly based on observations made with ARNICA@TNG)

**Claudi, R. U.; Bonanno, A.; Leccia, S.; Ventura, R.; Desidera, S.; Gratton, R.; Cosentino, R.; Paternò, L.; Endl, M.**, "Asteroseismology of Procyon A with SARG at TNG", 2005, A&A, 429, L17 (based on observations made with SARG@TNG)

**Costado, M. T.; Béjar, V. J. S.; Caballero, J. A.; Rebolo, R.; Acosta-Pulido, J.; Manchado, A.**, "A search for planetary-mass objects and brown dwarfs in the Upper Scorpius association", 2005, A&A, 443, 1021 (partly based on observations made with OIG@TNG)

**Davies, J. I.; Roberts, S.; Sabatini, S.**, "Searching for tidal tails – investigating galaxy harassment", 2005, MNRAS, 356, 794 (partly based on observations made with DOLoRes@TNG)

**De Propris, R.; Liske, J.; Driver, S. P.; Allen, P. D.; Cross, N. J. G.**, "The Millennium Galaxy Catalogue: Dynamically Close Pairs of Galaxies and the Global Merger Rate", 2005, AJ, 130, 1516 (partly based on observations made with DOLoRes@TNG)

**Della Ceca, R.; Ballo, L.; Braito, V.; Maccacaro, T.**, "The XMM-Newton view of the relativistic spectral features in AXJ0447-0627", 2005, ApJ, 627, 706 (partly based on observations made with DOLoRes@TNG)

**Driver, S. P.; Liske, J.; Cross, N. J. G.; De Propris, R.; Allen, P. D.**, "The Millennium Galaxy Catalogue: the space density and surface brightness distribution(s) of galaxies", 2005, MNRAS, 360, 81 (partly based on observations made with DOLoRes@TNG)

**Ellis, S. C.; Driver, S. P.; Allen, P. D.; Liske, J.; Bland-Hawthorn, J.; De Propris, R.**, "The Millennium Galaxy Catalogue: on the natural subdivision of galaxies", 2005, MNRAS, 363, 1275 (partly based on observations made with DOLoRes@TNG)

**Fiorenzano, A. F. M.; Gratton, R. G.; Desidera, S.; Cosentino, R.; Endl, M.**, "Line bisectors and radial velocity jitter from SARG spectra", 2005, A&A, 442, 775 (based on observations made with SARG@TNG)

**Galbiati, E.; Caccianiga, A.; Maccacaro, T.; Braito, V.; Della Ceca, R.; Severgnini, P.; Brunner, H.; Lehmann, I.; Page, M. J.**, "XMM-Newton spectroscopy of an X-ray selected sample of RL AGNs", 2005, A&A, 430, 927 (partly based on observations made with DOLoRes@TNG)

**Gänsicke, B. T.; Marsh, T. R.; Edge, A.; Rodriguez-Gil, P.; Steeghs, D.; et al.**, "Cataclysmic variables from a ROSAT/2MASS selection – I. Four new intermediate polars", 2005, MNRAS, 361, 141 (partly based on observations made with DOLoRes@TNG)

**Hendry, M. A.; Smartt, S. J.; Maund, J. R.; Pastorello, A.; et al.**, "A Study of the Type II-P Supernova 2003gd in M74", 2005, MNRAS, 359, 906 (partly based on observations made with DOLoRes@TNG)

**Kotak, R.; Meikle, W. P. S.; Pignata, G.; Stehle, M.; et al.**, "Spectroscopy of the type Ia supernova SN 2002er: days -11 to +215", 2005, A&A, 436, 1021 (partly based on observations made with DOLoRes@TNG)

**Lazzarin, M.; Marchi, S.; Magrin, S.; Licandro, J.**, "Spectroscopic investigation of near-Earth objects at Telescopio Nazionale Galileo", 2005, MNRAS, 359, 1575 (based on observations made with DOLoRes and NICS@TNG)

**Leipski, C.; Haas, M.; Meusinger, H.; Siebenmorgen, R.; Chini, R.; Scheyda, C. M.; et al.**, "The ISO-2MASS AGN survey: on the type-1 sources", 2005, A&A, 440, L5 (partly based on observations made with DOLoRes@TNG)

**Licandro, J.; Pinilla Alonso, N.**, "The Inhomogeneous Surface of Centaur 32522 Thereus (2001 $PT_{13}$)", 2005, ApJ, 630, 93L (based on observations made with NICS@TNG)

**Lodieu, N.**, "A study of the young open cluster Collinder 359", 2005, Astr. Nachr., 326, 1001 (partly based on observations made with DOLoRes@TNG)

**Longhetti, M.; Saracco, P.; Severgnini, P.; Della Ceca, R.; Braito, V.; Mannucci, F.; Bender, R.; Drory, N.; Feulner, G.; Hopp, U.**, "Dating the stellar population in massive early-type galaxies at z ~ 1.5", 2005, MNRAS, 361, 897 (based on observations made on NICS@TNG)

**Lucatello, S.; Tsangarides, S.; Beers, T. C.; Carretta, E.; Gratton, R. G.; Ryan, S. G.**, "The Binary Frequency Among Carbon-enhanced, s-Process-rich, Metal-poor

Stars", 2005, ApJ, 625, 825 (partly based on observations made with SARG@TNG)

**Marchesini, D.; Capetti, A.; Celotti, A.**, "Probing the nuclear obscuration in radio-galaxies with near infrared imaging", 2005, A&A, 433, 841 (based on observations made with ARNICA and NICS@TNG)

**Masetti, N.; Palazzi, E.; Pian, E.; Hunt, L.; Fynbo, J. P. U.; et al.**, "Late-epoch optical and near-infrared observations of the GRB 000911 afterglow and its host galaxy", 2005, A&A, 438, 841 (partly based on observations made with DOLoRes@TNG)

**Meusinger, H.; Froebrich, D.; Haas, M.; Irwin, M.; Laget, M.; Scholz, R.-D.**, " VPMS J1342+2840 - an unusual quasar from the variability and proper motion survey", 2005, A&A, 433, L25 (partly based on observations made with DOLoRes@TNG)

**Munari, U.; Henden, A.; Vallenari, A.; Bond, H. E.; Corradi, R. L. M.; Crause, L.; Desidera, S.; et al.**, "On the distance, reddening and progenitor of V838 Mon", 2005, A&A, 434, 1107 (partly based on observations made with SARG@TNG)

**Neiner, C.; Hubert, A.-M.; Catala, C.**, "The Identification of New Be Stars in GAUDI", 2005, ApJS, 156, 237 (partly based on observations made with SARG@TNG)

**Ortolani, S.; Bica, E.; Barbuy, B.**, "Color-Magnitude analysis of five old open clusters", 2005, A&A, 437, 531 (based on observations made with OIG@TNG)

**Ortolani, S.; Bica, E.; Barbuy, B.; Zoccali, M.**, "The old open clusters Berkeley 36, Berkeley 73 and Biurakan 13 (Berkeley 34)", 2005, A&A, 429, 607 (based on observations made with OIG@TNG)

**Poretti, E.; Alonso, R.; Amado, P. J.; Belmonte, J. A.; Garrido, R.; Martín-Ruiz, S.; Uytterhoeven, K.; et al.**, "Preparing the COROT Space Mission: New Variable Stars in the Galactic Anticenter Direction", 2005, AJ, 129, 2461 (partly based on observations made with SARG@TNG)

**Raiteri, C. M.; Villata, M.; Ibrahimov, M. A.; Larionov, V. M.; Kadler, M.**, "The WEBT campaign to observe AO 0235+16 in the 2003-2004 observing season. Results from radio-to-optical monitoring and XMM-Newton observations", 2005, A&A, 438, 39 (partly based on observations made with NICS@TNG)

**Ryabchikova, T.; Leone, F.; Kochukhov, O.**, "Abundances and chemical stratification analysis in the atmosphere of Cr-type Ap star HD 204411", 2005, A&A, 438, 973 (based on observations made with SARG@TNG)

**Santos, N. C.; Israelian, G.; Mayor, M.; Bento, J. P.; Almeida, P. C.; Sousa, S. G.; Ecuvillon, A.**, "Spectroscopic metallicities for planet-host stars: extending the samples", 2005, A&A, 437, 1127 (partly based on observations made with SARG@TNG)

**Saracco, P.; Longhetti, M.; Severgnini, P.; Della Ceca, R.; Braito, V.; Mannucci, F.; Bender, R.; Drory, N.; Feulner, G.; Hopp, U.; Maraston, C.**, "The density of very massive evolved galaxies to z ~ 1.7", 2005, MNRAS, 357, L40 (based on observations made with NICS@TNG)

**Scholz, R.-D.; Meusinger, H.; Jahreiß, H.**, "Search for nearby stars among proper motion stars selected by optical-to-infrared photometry. III. Spectroscopic distances of 322 NLTT stars", 2005, A&A, 442, 211 (partly based on observations made with DOLoRes@TNG)

**Severgnini, P.; Della Ceca, R.; Braito, V.; Saracco, P.; et al.**, "Looking for obscured QSOs in the X-ray emitting ERO population", 2005, A&A, 431, 87 (based on observations made with NICS@TNG)

**Solano, E.; Catala, C.; Garrido, R.; Poretti, E.; Janot-Pacheco, E.; Gutiérrez, R.; González, R.; et al.**, "GAUDI: A Preparatory Archive for the COROT Mission", 2005, AJ, 129, 547 (partly based on observations made with SARG@TNG)

**Tagliaferri, G.; Antonelli, L. A.; Chincarini, G.; Fernández-Soto, A.; Malesani, D.; Della Valle, M.; et al.**, "GRB 050904 at redshift 6.3: observations of the oldest cosmic explosion after the Big Bang", 2005, A&A, 443, L1 (partly based on observations made with NICS@TNG)

**Wolter, A.; Gioia, I. M.; Henry, J. P.; Mullis, C. R.**, "Unobscured QSO 2: a new class of objects?", 2005, A&A, 444, 165 (partly based on observations made with DOLoRes@TNG)

**Zappacosta, L.; Maiolino, R.; Finoguenov, A.; Mannucci, F.; Gilli, R.; Ferrara, A.**, "Constraining the thermal history of the Warm-Hot Intergalactic Medium", 2005, A&A, 434, 801 (partly based on observations made with DOLoRes@TNG)

# Publications 2005 on international non-refereed journals

**Arnaboldi, M.**, "Diffuse light in clusters of galaxies", 2005, Proceedings of "Baryons in dark matter halos", meeting held in Novigrad (Croazia), October 5-9 2004. Edited by Dettmar, R. J.; Klein, U.; Salucci, P.

**Barbieri, C.; Cremonese, G.; Verani, S.; Cosentino, R.; Leblanc, F.; Mendillo, M.; Sprague, A.; Hunten, D.**, "Observations of Mercury's Na-D emission spectrum with the TNG in August 2003", 2005, 35th COSPAR Scientific Assembly. Held 18 - 25 July 2004, in Paris, France, p. 2440

**Caballero, J. A.; Bejar, V. J. S.**, "Direct Detection of Giant Exoplanets", 2005, ING Newsletter, 9, 11

**Campins, H.; Licandro, J.; Ziffer, J.; Fernandez, Y. R.; Hora, J.; Kassis, M.; Pinilla Alonso, N.**, "Surface Characteristics of Comet-Asteroid Transition Objects 944 Hidalgo and 162P/Siding Spring (2004 TU12)", 2005, American Astronomical Society, DPS meeting #37, #16.02

**Capria, M. T.; Cremonese, G.; De Sanctis, M. C.**, "Forbidden oxygen lines in the spectrum of 153P/2002 C1 (Ikeya-Zhang)", 2005, Mem. SAIt Suppl., 6, 98

**Capria, M. T.; Cremonese, G.; De Sanctis, M. C.**, "High spectral resolution catalogue of comet 153P/2002 C1 Ikeya Zhang", 2005, American Astronomical Society, DPS meeting #37, #16.03

**Capria, M. T.; Cremonese, G.; Kawakita, H.; Watanabe, J.; de Sanctis, M. C.**, "Analysis of the high resolution spectrum of comet C/2002 C1 Ikeya-Zhang", 2005, 35th COSPAR Scientific Assembly. Held 18 - 25 July 2004, in Paris, France, p. 2488

**Covino, S.; Israel, G. L.; Antonelli, L. A.; Malesani, D.; Melandri, A.; Masetti, N.; Tagliaferri, G.**, "GRB050509B: Optical/Infared observations", 2005, GRB Coordinates Network, Circular Service, 3413, 1

**Covino, S.; Piranomonte, S.; Malesani, D.; Antonelli, L. A.; Boschin, W.; Mainella, G.**, "GRB 051215: near-infrared observations", 2005, GRB Coordinates Network, Circular Service, 4354, 1

**Cremonese, G.; Capria, M. T.; Kawakita, H.; Watanabe, J.**, "Catalog of emission lines of comet 2P/Encke", 2005, 35th COSPAR Scientific Assembly. Held 18 - 25

July 2004, in Paris, France, p.2287

D'Avanzo, P.; Fugazza, D.; Covino, S.; Malesani, D.; Masetti, N.; et al., "GRB 050306: optical afterglow", 2005, GRB Coordinates Network, Circular Service, 3089, 1

D'Avanzo, P.; Fugazza, D.; Masetti, N.; Pedani, M., "GRB050401: TNG r-band observation", 2005, GRB Coordinates Network, Circular Service, 3171, 1

de Leon Cruz, J.; Licandro, J.; Duffard, R.; Serra-Ricart, M., "Spectral and mineralogical characterization of NEOs", 2005, 35th COSPAR Scientific Assembly. Held 18 - 25 July 2004, in Paris, France, p.1379

Dotto, E.; Fornasier, S.; Barucci, M. A.; Licandro, J.; Boehnhardt, H.; Hainaut, O.; Marzari, F.; De Bergh, C.; De Luise, F., "Jupiter Trojans: a Survey of Members of Dynamical Families", 2005, American Astronomical Society, DPS meeting #37, #03.05

Duffard, R.; Lazzaro, D.; Licandro, J.; Desanctis, M. C.; Capria, M. T.; Carvano, J., "Vestoids and V-type asteroids: A Mineralogical Characterization", 2005, 35th COSPAR Scientific Assembly. Held 18 - 25 July 2004, in Paris, France, p. 408

Elias de la Rosa, N.; Benetti, S.; Cappellaro, E.; Dolci, M.; Pastorello, A.; Boschin, W.; Pinilla Alonso, N., "Possible Supernova in NGC 4656", 2005, IAUC, 8498, 1

Fugazza, D.; Antonelli, L. A.; Testa, V.; Di Fabrizio, L.; Tessicini, G., "GRB 050421: NIR observations", 2005, GRB Coordinates Network, Circular Service, 3300, 1

Fugazza, D.; D'Avanzo, P.; Covino, S.; Malesani, D.; et al., "GRB 050306: R-band observations", 2005, GRB Coordinates Network, Circular Service, 3078, 1

Israel, G. L.; dall'Osso, S.; Mangano, V.; Stella, L.; Covino, S.; Fugazza, D.; Campana, S.; Marconi, G.; Mereghetti, S.; Munari, U., "RX J0806.3-1527: Ten Years of Phase Coherent Monitoring in the Optical and X-ray Bands", 2005, proceedings of the IAP Conference "Interacting Binaries: Accretion, Evolution, and Outcomes", vol. 797, p. 307

**Lazzarin, M.; Magrin, S.; Marchi, S.**, "SINEO: Spectroscopic Investigation of Near Earth Object", 2005, Mem. SAIt Suppl., 6, 92

**Lazzarin, M.; Marchi, S.; Magrin, S.; Barucci, M. A.; di Martino, M.; Barbieri, C.**, "Visible and Near-Infrared Spectroscopic Investigation of Near-Earth Objects", 2005, 35th COSPAR Scientific Assembly. Held 18 - 25 July 2004, in Paris, France, p. 2722

**Lazzarin, M.; Marchi, S.; Magrin, S.; Danese, S.**, "Ongoing results of SINEO (Spectroscopic Investigation of Near Earth Objects) survey", 2005, American Astronomical Society, DPS meeting #37, #07.09

**Licandro, J.; de León Cruz, J.; Serra-Ricart, M.; García, A.; Pinilla Alonso, N.**, "The physical nature of asteroids in cometary orbit", 2005, 35th COSPAR Scientific Assembly. Held 18 - 25 July 2004, in Paris, France, p. 3336

**Maiorano, E.; Palazzi, E.; Masetti, N.; Malesani, D.; D'Avanzo, P.; Israel, G. L.; Chincarini, G.; Stella, L.; Pedani, M.**, "GRB050712: TNG r-band observation", 2005, GRB Coordinates Network, Circular Service, 3601, 1

**Malesani, D.; Piranomonte, S.; Fiore, F.; Tagliaferri, G.; Fugazza, D.; Cosentino, R.**, "GRB 050525: observatons of the optical afterglow", 2005, GRB Coordinates Network, Circular Service, 3469, 1

**Mannucci, F.; Covino, S.; Malesani, D.**, "GRB 050820A: NIR photometry at the TNG", 2005, GRB Coordinates Network, Circular Service, 4045, 1

**Mendez, J.; Ruiz-Lapuente, P.; Altavilla, G.; Balastegui, A.; Irwin, M.; Schahmaneche, K.; Balland, C.; Pain, R.; Walton, N.**, "Supernova search at intermediate z. II. Host galaxy morphology", 2005, *1604-2004: Supernovae as Cosmological Lighthouses*, ASP Conference Proceedings Vol. 342, p. 488

**Pedani, M.**, "An Updated View of the Light Pollution at the Roque de Los Muchachos Observatory", 2005, ING Newsletter, 9, 28

**Pinilla Alonso, N.; Licandro, J.**, "Mineralogycal analysis of the icy surface of TNOs (50000) Quaoar and 2002 $TX_{300}$", 2005, 35th COSPAR Scientific Assembly. Held 18 - 25 July 2004, in Paris, France, p. 3330

**Piranomonte, S.; Covino, S.; Antonelli, A.; Malesani, D.**, "GRB 051105A: TNG

optical observations", 2005, GRB Coordinates Network, Circular Service, 4201, 1

**Piranomonte, S.; Magazzu, A.; Mainella, G.; Fiore, F.; Covino, S.; Fugazza, D.; et al.**, "GRB050922C: TNG spectroscopic observations", 2005, GRB Coordinates Network, Circular Service, 4032, 1

**Saba, L.; Capria, M. T.; Cremonese, G.**, "High-resolution observations of 2/P Encke comet: preliminary results", 2005, Mem. SAIt Suppl., 6, 137

**Testa, V.; Melandri, A.; Antonelli, L. A.; Stella, L.; Covino, S.; Fugazza, D.; Tagliaferri, G.; et al.**, "GRB050802: TNG optical observations", 2005, GRB Coordinates Network, Circular Service, 3765, 1

**Vernazza, P.; Fulchignoni, M.; Birlan, M.**, "Spectroscopic investigation of Near-Earth Objects", 2005, 35th COSPAR Scientific Assembly. Held 18 - 25 July 2004, in Paris, France, p. 2456

**Zampieri, L.; Ramina, M.; Pastorello, A.**, "Understanding Type II Supernovae", 2005, *Cosmic Explosions, On the 10th Anniversary of SN1993J*. Proceedings of IAU Colloquium 192. Edited by J.M. Marcaide and Kurt W. Weiler. Springer Proceedings in Physics, vol. 99. Berlin: Springer, p. 275

# Publications 2004 on international journals with "referee"

**Anandarao, B. G.; Chakraborty, A.; Ojha, D. K.; Testi, L.**, "Detection of knots and jets in IRAS 06061+2151", 2004, A&A, 421, 1045 (partly based on observations made with NICS@TNG)

**Andreuzzi, G.; Bragaglia, A.; Tosi, M.; Marconi, G.**, "UBVI photometry of the intermediate-age open cluster NGC 6939", 2004, MNRAS, 348, 297 (based on observations made with DOLoRes@TNG)

**Barbieri, C.; Verani, S.; Cremonese, G.; Sprague, A.; Mendillo, M.; Cosentino, R.; Hunten, D.**, "First observations of the Na exosphere of Mercury with the high-resolution spectrograph of the 3.58 m Telescopio Nazionale Galileo", 2004, P&SS, 52, 1169 (based on observations made with SARG@TNG)

**Battinelli, P.; Demers, S.**, "Carbon star survey in the Local Group VIII. Probing the stellar halo of NGC 147", 2004, A&A, 418, 33 (partly based on observations made with DOLoRes@TNG).

**Bellazzini, M.; Gennari, N.; Ferraro, F. R.; Sollima, A.**, "The distance to the Leo I dwarf spheroidal galaxy from the red giant branch tip", 2004, MNRAS, 354, 708 (based on observations made with DOLoRes@TNG)

**Boschin, W.; Girardi, M.; Barrena, R.; Biviano, A.; Feretti, L.; Ramella, M.**, "Internal dynamics of the radio-halo cluster A2219: A multi-wavelength analysis", 2004, A&A, 416, 839 (partly based on observations made with DOLoRes@TNG).

**Coccato, L.; Corsini, E. M.; Pizzella, A.; Morelli, L.; Funes, J. G.; Bertola, F.**, "Minor-axis velocity gradients in disk galaxies", 2004, A&A, 416, 507 (partly based on observations made with DOLoRes@TNG)

**Covino, E.; Frasca, A.; Alcalá, J. M.; Paladino, R.; Sterzik, M. F.**, "Improved fundamental parameters for the low-mass pre-main sequence eclipsing system RX J0529.4+0041", 2004, A&A, 427, 637 (partly based on observations made with OIG@TNG)

**de Leon, J.; Duffard, R.; Licandro, J.; Lazzaro, D.**, "Mineralogical characterization of A-type asteroid (1951) Lick", 2004, A&A, 422, 59 (based on observations made with NICS@TNG)

**de Wit, W. J.; Testi, L.; Palla, F.; Vanzi, L.; Zinnecker, H.**, "The origin of massive O-type field stars. I. A search for clusters", 2004, A&A, 425, 937 (partly based on observations made with NICS@TNG)

**Della Ceca, R.; Maccacaro, T.; Caccianiga, A.; Severgnini, P.; Braito, V.; Barcons, X.; Carrera, F. J.; Watson, M. G.; Tedds, J. A.; Brunner, H.; et al.**, "Exploring the X-ray sky with the XMM-Newton bright serendipitous survey", 2004, 428, 383 (partly based on observations made with DOLoRes@TNG).

**Desidera, S.; Gratton, R. G.; Endl, M.; Claudi, R. U.; Cosentino, R.**, "No planet around HD 219542 B", 2004, A&A, 420, L27 (based on observations made with SARG@TNG)

**Desidera, S.; Gratton, R.; Scuderi, S.; Claudi, R.; Cosentino, R.; Barbieri, M.; Bonanno, G.; Carretta, E.; Endl, M.; Lucatello, S.; Martinez-Fiorenzano, A.; Marzari, F.**, "Abundance differences between components of wide binaries", 2004, A&A, 420, 683 (based on observations made with SARG@TNG)

**Duffard, R.; Lazzaro, D.; Licandro, J.; de Sanctis, M. C.; Capria, M. T.; Carvano, J. M.**, "Mineralogical characterization of some basaltic asteroids in the neighborhood of (4) Vesta: first results", 2004, Icarus, 171, 120 (based on observations made with NICS@TNG)

**Ecuvillon, A.; Israelian, G.; Santos, N. C.; Mayor, M.; García López, R. J.; Randich, S.**, "Nitrogen abundances in planet-harbouring stars", 2004, A&A, 418, 703 (partly based on observations made with SARG@TNG)

**Ecuvillon, A.; Israelian, G.; Santos, N. C.; Mayor, M.; Villar, V.; Bihain, G.**, "C, S, Zn and Cu abundances in planet-harbouring stars", 2004, A&A, 426, 619 (partly based on observation made with SARG@TNG)

**Fontani, F.; Cesaroni, R.; Testi, L.; Molinari, S.; Zhang, Q.; Brand, J.; Walmsley, C. M.**, "Nature of two massive protostellar candidates: IRAS 21307+5049 and IRAS 22172+5549", 2004, A&A, 424, 179 (partly based on observations made with NICS@TNG)

**Fornasier, S.; Dotto, E.; Barucci; M. A.; Barbieri, C.**, "Water ice on the surface of the large TNO 2004 DW", 2004, A&A, 422, L43 (based on observations made with NICS@TNG)

**Fulle, M.; Barbieri, C.; Cremonese, G.; Rauer, H.; Weiler, M.; Milani, G.;**

**Ligustri, R.**, "The dust environment of comet 67P/Churyumov-Gerasimenko", 2004, A&A, 422, 357 (partly based on observations made with DOLoRes@TNG)

**Galleti, S.; Bellazzini, M.; Ferraro, F. R.**, "The distance of M 33 and the stellar population in its outskirts", 2004, A&A, 423, 925 (based on observations made with DOLoRes@TNG)

**Grazian, A.; Negrello, M.; Moscardini, L.; Cristiani, S.; Haehnelt, M. G.; Matarrese, S.; Omizzolo, A.; Vanzella, E.**, "The Asiago-ESO/RASS QSO Survey. III. Clustering Analysis and Theoretical Interpretation", 2004, AJ, 127, 592 (partly based on observations made with DOLoRes@TNG)

**Hubrig, S.; Kurtz, D. W.; Bagnulo, S.; Szeifert, T.; Schöller, M.; Mathys, G.; Dziembowski, W. A.**, "Measurements of magnetic fields over the pulsation cycle in six roAp stars with FORS 1 at the VLT", 2004, A&A, 415, 661 (partly based on observations made with SARG@TNG)

**Israelian, G.; Santos, N. C.; Mayor, M.; Rebolo, R.**, "Lithium in stars with exoplanets", 2004, A&A, 414, 601 (partly based on observations made with SARG@TNG)

**Israelian, G.; Shchukina, N.; Rebolo, R.; Basri, G.; Gonzalez Hernandez, J. I.; Kajino, T.**, "Oxygen and magnesium abundance in the ultra-metal-poor giants CS 22949-037 and CS 29498-043: Challenges in models of atmosphere", 2004, A&A, 419, 1095 (partly based on observations made with SARG@TNG)

**Kawakita, H.; Watanabe, J.; Furusho, R.; Fuse, T.; Capria, M. T.; De Sanctis, M. C.; Cremonese, G.**, "Spin Temperatures of Ammonia and Water Molecules in Comets", 2004, ApJ, 601, 1152 (partly based on observations made with SARG@TNG).

**Klose, S.; Greiner, J.; Rau, A.; Henden, A. A.; Hartmann, D. H.; Zeh, A.; Ries, C.; Masetti, N.; Malesani, D.; Guenther, E.; et al.**, "Probing a Gamma-Ray Burst Progenitor at a Redshift of z = 2: A Comprehensive Observing Campaign of the Afterglow of GRB 030226", 2004, AJ, 128, 1942 (partly based on observations made with DOLoRes@TNG)

**Leone, F.; Catanzaro, G.**, "The effect of the surface distribution of elements on measuring the magnetic field of chemically peculiar stars. The case of the roAp star HD 24712", 2004, A&A, 425, 271 (partly based on observations made with SARG@TNG)

**Maier, C.; Meisenheimer, K.; Hippelein, H.**, "The metallicity-luminosity relation at medium redshift based on faint CADIS emission line galaxies", 2004, A&A, 418, 475 (partly based on observations made with DOLoRes@TNG)

**Maiolino, R.; Oliva, E.; Ghinassi, F.; Pedani, M.; Mannucci, F.; Mujica, R.; Juarez, Y.**, "Extreme gas properties in the most distant quasars", 2004, A&A, 420, 889 (based on observations made with NICS@TNG)

**Maiolino, R.; Schneider, R.; Oliva, E.; Bianchi, S.; Ferrara, A.; Mannucci, F.; Pedani, M.; Roca Sogorb, M.**, "A supernova origin for dust in a high-redshift quasar", 2004, Nature, 431, 533 (based on observations made with NICS@TNG)

**Massi, F.; Codella, C.; Brand, J.**, "Discovery of [FeII]- and H2-emission from protostellar jets in the CB3 and CB230 globules", 2004, A&A, 419, 241 (based on observations made with NICS@TNG)

**Moreno, F.; Lara, L. M.; Muñoz, O.; López-Moreno, J. J.; Molina, A.**, "Dust in Comet 67P/Churyumov-Gerasimenko", 2004, ApJ, 613, 1263 (based on observations made with DOLoRes@TNG).

**Moretti, A.; Guzzo, L.; Campana, S.; Lazzati, D.; Panzera, M. R.; Tagliaferri, G.; Arena, S.; Braglia, F.; Dell'Antonio, I.; Longhetti, M.**, "The Brera Multi-scale Wavelet HRI Cluster Survey. I. Selection of the sample and number counts", 2004, A&A, 428, 21 (partly based on observations made with DOLoRes@TNG)

**Netzer, H.; Shemmer, O.; Maiolino, R.; Oliva, E.; Croom, S.; Corbett, E.; Di Fabrizio, L.**, "Near-Infrared Spectroscopy of High-Redshift Active Galactic Nuclei. II. Disappearing Narrow-Line Regions and the Role of Accretion", 2004, ApJ, 614, 558 (based on observations made with NICS@TNG)

**Ojha, D. K.; Ghosh, S. K.; Kulkarni, V. K.; Testi, L.; Verma, R. P.; Vig, S.**, "A study of the Galactic star forming region IRAS 02593+6016/S 201 in infrared and radio wavelengths", 2004, A&A, 415, 1039 (based on observations made with NICS@TNG)

**Origlia, L.; Ranalli, P.; Comastri, A.; Maiolino, R.**, "Stellar and Gaseous Abundances in M82", 2004, ApJ, 606, 862 (based on observations made with NICS@TNG)

**Pedani, M.**, "Light pollution at the Roque de los Muchachos Observatory", 2004,

New Astronomy, 9, 641 (based on observations made with DOLoRes@TNG)

**Peroux, C.; Deharveng, J.-M.; Le Brun, V.; Cristiani, S.**, "Classical and MgII-selected damped Lyman-alpha absorbers: impact on $\Omega_{HI}$ at z < 1.7", 2004, MNRAS, 352, 1291 (based on observations made with DOLoRes@TNG)

**Pignata, G.; Patat, F.; Benetti, S.; Blinnikov, S.; Hillebrandt, W.; Kotak, R.; Leibundgut, B.; Mazzali, P. A.; Meikle, P.; Qiu, Y.; et al.**, "Photometric observations of the Type Ia SN 2002er in UGC 10743", 2004, MNRAS, 355, 178 (partly based on observations made with DOLoRes@TNG)

**Sabbadin, F.; Turatto, M.; Cappellaro, E.; Benetti, S.; Ragazzoni, R.**, "The 3-D ionization structure and evolution of NGC 7009 (Saturn Nebula)", 2004, A&A, 416, 955 (partly based on observations made with SARG@TNG)

**Santos, N. C.; Israelian, G.; Mayor, M.**, "Spectroscopic [Fe/H] for 98 extra-solar planet-host stars. Exploring the probability of planet formation", 2004, A&A, 415, 1153 (partly based on observations made with SARG@TNG)

**Sestito, P.; Randich, S.; Pallavicini, R.**, "Lithium evolution in intermediate age and old open clusters: NGC 752 revisited", 2004, A&A, 426, 809 (based on observations made with SARG@TNG)

**Shemmer, O.; Netzer, H.; Maiolino, R.; Oliva, E.; Croom, S.; Corbett, E.; Di Fabrizio, L.**, "Near-Infrared Spectroscopy of High-Redshift Active Galactic Nuclei. I. A Metallicity-Accretion Rate Relationship", 2004, ApJ, 614, 547 (partly based on observations made with NICS@TNG)

**Valenti, E.; Ferraro, F. R.; Perina, S.; Origlia, L.**, "Near-IR photometry of five Galactic Globular Clusters", 2004, A&A, 419, 139 (based on observations made with ARNICA@TNG)

**Villata, M.; Raiteri, C. M.; Kurtanidze, O. M.; Nikolashvili, M. G.; Ibrahimov, M. A.; Papadakis, I. E.; Tosti, G.; Hroch, F.; Takalo, L. O.; Sillanpää, A.; et al.**, "The WEBT Campaign 2001 and its extension. Optical light curves and colour analysis 1994-2002", 2004, A&A, 421, 103 (partly based on observations made with DOLoRes@TNG)

**Zapatero Osorio, M. R.; Martin, E. L.**, "A CCD imaging search for wide metal-poor binaries", 2004, A&A, 419, 167 (partly based on observations made with

## Publications 2004 on international non-refereed journals

**Aparicio, A.; Rosenberg, A.; Piotto, G.; Saviane, I.; Recio-Blanco, A.**, "The Milky Way Formation Timescale", 2004, Mem. SAIt, 75, 13

**Benetti, S.; Elias de la Rosa, N.; Blanc, G.; Navasardyan, H.; Turatto, M.; Zampieri, L.; Cappellaro, E.; Pedani, M.**, "Supernova 2004aw in NGC 3997", 2004, IAU Circular, 8312, 3

**Bettoni, D.; Mazzei, P.; Della Valle, A.; Franceschini; G., DeZotti, A.; Aussel, H.**, "Optical and Spectroscopic Follow-Up of the Deeper FIR Selected Sample", 2004, Mem. SAIt Suppl., 5, 267

**Binzel, R. P.; Licandro, J.; Serra-Ricart, M.; de Leon Cruz, J.; Pinilla Alonso, N.**, "Comet C/2003 WT42 (LINEAR)", 2004, IAU Circular, 8270, 1

**Bonanno, G.; Cosentino, R.; Belluso, M.; Bruno, P.; Bortoletto, F.; D'Alessandro, M.; Fantinel, D.; Giro, E.; Corcione, L.; Carbone, A.; Evola, G.**, "The New Generation CCD Controller: First Results", 2004, *Scientific Detectors for Astronomy, The Beginning of a New Era*, eds. P. Amico; J. W. Beletic; J. E. Beletic, p. 423.

**Cappellaro, E.**, "The evolution of the cosmic SN rate", 2004, Mem. SAIt, 75, 206

**Caproni, A.; Zacchei, A.; Vuerli, C.; Pucillo, M.**, "Porting and refurbishment of the WSS TNG control software", 2004, Proceedings of the SPIE, Volume 5496, pp. 538-546

**Claudi, R. U.; Barbieri, M.; Bonanno, G.; Carretta, E.; Cosentino, R.; Desidera, S.; Endl, M.; Gratton, R.; Lucatello, S.; Marzari, F.; Scuderi, S.**, "The SARG search for planets: first results", 2004, Second Eddington Workshop: *Stellar structure and habitable planet finding*, 9 - 11 April 2003, Palermo, Italy. Edited by F. Favata, S. Aigrain and A. Wilson. ESA SP-538, Noordwijk: ESA Publications Division, ISBN 92-9092-848-4, 2004, p. 297 # 300

**Claudi, R. U.; Bonanno, A.; Leccia, S.; Ventura, R.; Desidera, S.; Gratton, R.; Cosentino, R.; Paternò, L.; Endl, M.**, "Asteroseismology of Procyon A with SARG at TNG", 2004, Proceedings of the SOHO 14 / GONG 2004 Workshop (ESA SP-

559). *Helio- and Asteroseismology: Towards a Golden Future.* 12-16 July, 2004. New Haven, Connecticut, USA. Editor: D. Danesy., p.125

**Claudi, R. U.; Bonanno, A.; Ventura, R.; Bonanno, G.; Cosentino, R.; Desidera, S.; Gratton, R.; Scuderi, S.; Endl, M.**, 2004, "Asteroseismology of Procyon: Preliminary results from SARG", 2004, Communications in Asteroseismology, vol. 145, p. 53

**Cosentino, R.; Bruno, P.; Gonzalez, M.; Huertas, M.; Scuderi, S.**, "Instrument remote control project at TNG: SARG implementation", 2004, *UV and Gamma-Ray Space Telescope Systems.* Edited by G. Hasinger; M. J. L. Turner. Proceedings of the SPIE, Volume 5492, pp. 891-899.

**D'Avanzo, P.; Covino, S.; Antonelli, L. A.; Fugazza, D.; et al.**, "GRB041006: break in the light curve", 2004, GRB Coordinates Network, Circular Service, 2788, 1

**D'Avanzo, P.; Fugazza, D.; Melandri, A.; Malesani, D.; Tagliaferri, G.; Antonelli, L. A.; Campana, S.; Chincarini, G.; Covino, S.; Cucchiara, A.; et al.**, "GRB040624: optical monitoring at TNG", 2004, GRB Coordinates Network, Circular Service, 2632, 1

**Desidera, S.; Gratton, R. G.; Claudi, R. U.; Carretta, E.; Lucatello, S.; Martinez-Fiorenzano, A.; Bonanno, G.; Cosentino, R.; Scuderi, S.; Barbieri, M.; et al.**, "Searching for Planets around Stars in Wide Binaries", 2004, *Extrasolar Planets: Today and Tomorrow*, ASP Conference Proceedings, Vol. 321, held 30 June - 4 July 2003, Institut D'Astrophysique de Paris, France. Edited by J.-P. Beaulieu, A. Lecavelier des Etangs and C. Terquem. ISBN: 1-58381-183-4, p.103

**Dotto, E.; Fornasier, S.; Barucci, M. A.; Boehnhardt, H.; Hainaut, O.; Marzari, F.; Licandro, J.; de Bergh, C.**, "Visible and near-infrared spectroscopic survey of Jupiter Trojan asteroids: investigation of dynamical families", 2004, American Astronomical Society, DPS meeting #36, #32.06

**Fugazza, D.; D'Avanzo, P.; Tagliaferri, G.; Kalogerakos, S.; Campana, S.; Chincarini, G.; Covino, S.; Cucchiara, A.; Malesani, D.; Masetti, N.; et al.**, "GRB 040624: optical observations at TNG and VLT", 2004, GRB Coordinates Network, Circular Service, 2617, 1

**Fugazza, D.; Fiore, F.; Covino, S.; Antonelli, L. A.; D'Avanzo, P.; Cocchia, F.; Malesani, D.; Pian, E.; Stella, L.; Lorenzi, V.; Tessicini, G.**, "GRB041006: optical photometry and spectroscopy at TNG", 2004, GRB Coordinates Network, Circular Service, 2782, 1

**Fugazza, D.; Mantegazza, L.; Antonello, E.**, "SV 39: a strange variable star in the field of the Local Group galaxy IC 1613", 2004, Mem. SAIt, 75, 122

**Ghedina, A.; Gaessler, W.; Cecconi, M.; Ragazzoni, R.; Puglisi, A.T.; De Bonis, F.**, "Latest developments on the loop control system of AdOpt@TNG", 2004, ALT'03 International Conference on *Advanced Laser Technologies: Biomedical Optics*. Edited by R. K. Wang; J. C. Hebden; A. V. Priezzhev; V. V. Tuchin. Proceedings of the SPIE, Volume 5490, pp. 1347-1355

**Ghedina, A.; Gonzalez, M.; Lodi, M.; Cecconi, M.; Oliva, E.; Scuderi, S.; Cosentino, R.; Caproni, A.**, "The new active optics system of TNG", 2004, Proceedings of the SPIE, Volume 5489, pp. 997-1003

**Ghedina, A.; Pedani, M.; Guerra, J. C.; Zitelli, V.; Porceddu, I.**, "Three years of dust monitoring at the Galileo telescope", 2004, Proceedings of the SPIE, Volume 5489, pp. 227-234

**Gratton, R. G.; Carretta, E.; Claudi, R. U.; Desidera, S.; Lucatello, S.; Barbieri, M.; Bonanno, G.; Cosentino, R.; Scuderi, S.; Endl, M.; Marzari, F.**, "The SARG exo-planets search", 2004, Mem. SAIt, 75, 97

**Israel, G. L.; Covino, S.; Dall'Osso, S.; Fugazza, D.; Mouche, C. W.; Stella, L.; Campana, S.; Mangano, V.; Marconi, G.; Bagnulo, S.; Munari, U.**, "Taking the pulse of the shortest orbital period binary system RX J0806.3+1527", 2004, Mem. SAIt Suppl., 5, 148

**Kinman, T. D.; Bragaglia, A.; Cacciari, C.; Buzzoni, A.; Spagna, A.**, "The Vertical Structure of the Halo Rotation", 2004, Mem. SAIt, 75, 36

**Lazzarin, M.; Marchi, S.; Magrin, S.; Barbieri, C.**, "SINEO: Spectroscopic Investigation of Near Earth Objects", 2004, Mem. SAIt Suppl., 5, 21

**Lazzaro, D.; Duffard, R.; de Leon, J.; Licandro, J.**, "Basaltic asteroids and HED meteorites: searching for a genetic link using spectral parameters", 2004, American Astronomical Society, DPS meeting #36, #32.01

**Li Causi, G.; de Luca, M.; Vitali, F.; Lorenzetti, D.**, "GO-CART: the GOHSS Calibration and Reduction Tool", 2004, SPIE Meeting *Astronomical Telescopes and Instrumentation*, 22-24 June 2004 Glasgow, Scotland United Kingdom. Edited by H.

Lewis, and G. Raffi. SPIE Proceedings, Vol. 5496, p. 765.

**Longo, G.; Donalek, C.; Raiconi, G.; Staiano, A.; Tagliaferri, R.; Pasian, F.; Sessa, S.; Smareglia, R.; Volpicelli, A.**, "Artificial Intelligence Tools for Data Mining in Large Astronomical Databases", 2004, *Toward an International Virtual Observatory*, Proceedings of the ESO/ESA/NASA/NSF Conference held in Garching, Germany, 10-14 June 2002. Edited by P.J. Quinn, and K.M. Gorski. ESO Astrophysics Symposia. Berlin: Springer, 2004, p. 202.

**Oliva, E.**, "Telescopio Nazionale Galileo: a status report", 2004, Mem. SAIt, 75, 218

**Oliva, E.; Origlia, L.; Maiolino, R.; Gennari, S.; Biliotti, V.; Rossetti, E.; Baffa, C.; Leone, F.; Montegriffo, P.; Lolli, M.; et al.**, "GIANO: an ultrastable IR echelle spectrometer optimized for high-precision radial velocity measurements and for high-throughput low-resolution spectroscopy", 2004, *UV and Gamma-Ray Space Telescope Systems*. Edited by Hasinger, G.; Turner, M. J. L. Proceedings of the SPIE, Volume 5492, pp. 1274-1279

**Pasian, F.; Benacchio, L.; Smareglia, R.**, "The Prototype TNG Long-Term Archive and its Interactions with the Italian GRID Project: Toward an International Virtual Observatory", 2004, Proceedings of the ESO/ESA/NASA/NSF Conference held in Garching, Germany, 10-14 June 2002. Edited by P.J. Quinn, and K.M. Gorski. ESO Astrophysics Symposia. Berlin: Springer, 2004, p. 88.

**Peroux, C.; Deharveng, J.-M.; Le Brun, V.; Cristiani, S.**, "Classical and MgII-selected Damped Lyman-alpha Absorbers: impact on $\Omega_{HI}$ at z < 1.7", 2004, SF2A-2004: Semaine de l'Astrophysique Francaise, meeting held in Paris, France, June 14-18, 2004, Eds.: F. Combes, D. Barret, T. Contini, F. Meynadier and L. Pagani EdP-Sciences

**Pinilla Alonso, N.; Licandro, J.; Campins, H.**, "Mineralogical analysis of two different kind of icy surfaces in the trans-neptunian belt, TNOs (50000) Quaoar and 2002 $TX_{300}$", 2004, American Astronomical Society, DPS meeting #36, #11.07

**Sestito, P.; Randich, S.; Pallavicini, R.**, "High resolution spectroscopy of open clusters with SARG", 2004, Mem. SAIt, 75, 24

**Spagna, A.; Lattanzi, M. G.; McLean, B.; Bucciarelli, B.; Carollo, D.; Drimmel, R.; et al.**, "Data mining with the multicolor GSC-II database", 2004, Mem. SAIt Suppl., 5, 97

# Publications 2003 on international journals with "referee"

**Aguerri, J. A. L.; Debattista, V. P.; Corsini, E. M.**, "Measurement of fast bars in a sample of early- type barred galaxies", 2003, MNRAS, 338, 465 (partly based on observations made with DOLoRes@TNG)

**Arnaboldi, M.; Freeman, K. C.; Okamura, S.; Yasuda, N.; Gerhard, O.; Napolitano, N. R.; Pannella, M.; Ando, H.; Doi, M.; Furusawa, H.; et al.**, "Narrowband Imaging in [O III] and Halfa to Search for Intracluster Planetary Nebulae in the Virgo Cluster", 2003, AJ, 125, 514 (based on observations made with DOLoRes@TNG)

**Bodaghee, A.; Santos, N. C.; Israelian, G.; Mayor, M.**, "Chemical abundances of planet-host stars. Results for α and Fe-group elements", 2003, A&A, 404, 715 (partly based on observations made with SARG@TNG)

**Carollo, D.; Koester, D.; Spagna, A.; Lattanzi, M. G.; Hodgkin, S. T.**, "Model atmosphere analysis of the extreme DQ white dwarf GSC2U J131147.2+292348", 2003, A&A, 400, L13 (based on observations with DOLoRes@TNG)

**Castro-Tirado, A. J.; Gorosabel, J.; Guziy, S.; Reverte, D.; Castro Cerón, J. M.; de Ugarte Postigo, A.; Tanvir, N.; Mereghetti, S.; Tiengo, A.; Buckle, J.; et al.**, " GRB 030227: The first multiwavelength afterglow of an INTEGRAL GRB", 2003, A&A, 411, L315 (partly based on observations with NICS@TNG)

**Catanzaro, G.; André, M. K.; Leone, F.; Sonnentrucker, P.**, "High resolution spectroscopy of HD207538 from Far-UV (FUSE) to Visible (SARG-TNG). A global picture of the stellar and interstellar features modeled", 2003, A&A, 404, 677 (partly based on observations made with SARG@TNG)

**Cellino, A.; Diolaiti, E.; Ragazzoni, R.; Hestroffer, D.; Tanga, P.; Ghedina, A.**, "Speckle interferometry observations of asteroids at TNG", 2003, Icarus, 162, 278 (based on observations made with the SPECKLE CAMERA@TNG)

**Christlieb, N.**, "Finding the Most Metal-poor Stars of the Galactic Halo with the Hamburg/ESO Objective-prism Survey", 2003, Rev. Mod. Astr., 16, 191 (partly based on observations made with SARG@TNG)

**Corsini, E. M.; Debattista, V. P.; Aguerri, J. A. L.**, "Direct Confirmation of Two Pattern Speeds in the Double Barred Galaxy NGC 2950", 2003, ApJ, 599, L29 (partly based on observations made with DOLoRes@TNG)

**Covino, S.; Malesani, D.; Tavecchio, F.; Antonelli, L. A.; Arkharov, A.; Di Paola, A.; Fugazza, D.; Ghisellini, G.; Larionov, V.; Lazzati, D.; et al.**, "Optical and NIR Observations of the Afterglow of GRB 020813", 2003, A&A, 404, L5 (partly based on observations made with DOLoRes and NICS@TNG)

**Desidera, S.; Gratton, R. G.; Endl, M.; Barbieri, M.; Claudi, R. U.; Cosentino, R.; Lucatello, S.; Marzari, F.; Scuderi, S.**, "A search for planets in the metal-enriched binary HD 219542", 2003, A&A, 405, 207 (based on observations made with SARG@TNG)

**Fiore, F.; Brusa, M.; Cocchia, F.; Baldi, A.; Carangelo, N.; Ciliegi, P.; Comastri, A.; La Franca, F.; Maiolino, R.; Matt, G.; et al.**, "The HELLAS2XMM survey. IV. Optical identifications and the evolution of the accretion luminosity in the Universe", 2003, A&A, 409, 79 (partly based on observations made with DOLoRes@TNG)

**Fornasier, S.; Barucci, M. A.; Binzel, R. P.; Birlan, M.; Fulchignoni, M.; Barbieri, C.; Bus, S. J.; Harris, A. W.; Rivkin, A. S.; Lazzarin, M.; et al.**, "A portrait of 4979 Otawara, target of the Rosetta space mission", 2003, A&A, 398, 327 (partly based on observations made with OIG@TNG)

**Gratton, R. G.; Carretta, E.; Claudi, R.; Lucatello, S.; Barbieri, M.**, "Abundances for metal-poor stars with accurate parallaxes. I. Basic data", 2003, A&A, 404, 187 (partly based on observations made with SARG@TNG)

**Gratton, R. G.; Carretta, E.; Desidera, S.; Lucatello, S.; Mazzei, P.; Barbieri, M.**, "Abundances for metal-poor stars with accurate parallaxes. II. $\alpha$-elements in the halo", 2003, A&A, 406, 131 (partly based on observations made with SARG@TNG).

**Israelian, G.; Santos, N. C.; Mayor, M.; Rebolo, R.**, "New measurement of the 6Li/7Li isotopic ratio in the extra-solar planet host star HD 82943 and line blending in the Li 6708 Å region", 2003, A&A, 405, 573 (partly based on observations made with SARG@TNG)

**Lara, L.-M.; Licandro, J.; Tozzi, G. P.**, "Dust in comet McNaught-Hartley (C/1999 T1) from Jan. 25 to Feb. 04, 2001: IR and optical CCD imaging", 2003, A&A, 404, 373 (partly based on observations made with DOLoRes and NICS@TNG)

**Lazzati, D.; Covino, S.; di Serego Alighieri, S.; Ghisellini, G.; Vernet, J.; Le Floc'h, E.; Fugazza, D.; Di Tomaso, S.; Malesani, D.; Masetti, N.; et al.**, "Intrinsic and dust-induced polarization in gamma- ray burst afterglows: The case of GRB

021004", 2003, A&A, 410, 823 (partly based on observations made with NICS@TNG)

**Leone, F.; Kurtz, D. W.**, "Discovery of magnetic field variations with the 12.1-minute pulsation period of the roAp star gamma Equulei", 2003, A&A, 407, L67 (based on observations made with SARG@TNG)

**Leone, F.; Vacca, W. D.; Stift, M. J.**, "Measuring stellar magnetic fields from high resolution spectroscopy of near-infrared lines", 2003, A&A, 409, 1055 (partly based on observations made with SARG@TNG)

**Licandro, J.; Campins, H.; Hergenrother, C.; Lara, L. M.**, "Near-infrared spectroscopy of the nucleus of comet 124P/Mrkos", 2003, A&A, 398, 45 (based on observations made with NICS@TNG)

**Lopez-Santiago, J.; Montes, D.; Fernandez- Figueroa, M. J.; Ramsey, L. W.**, "Rotational modulation of the photospheric and chromospheric activity in the young, single K2- dwarf PW And", 2003, A&A, 411, 489 (partly based on observations made with SARG@TNG)

**Maiolino, R.; Juarez, J.; Mujica, R.; Nagar, N.; Oliva, E.**, "Early star formation traced by the highest redshift quasars", 2003, ApJ, 596, L155 (based on observations made with NICS@TNG)

**Mannucci, F.; Maiolino, R.; Cresci, G.; Della Valle, M.; Vanzi, L.; Ghinassi, F.; Ivanov, V. D.; Nagar, N. M.; Alonso-Herrero, A.**, "The infrared supernova rate in starburst galaxies", 2003, A&A, 401, 519 (based on observations made with NICS@TNG)

**Marconi, A.; Axon, D. J.; Capetti, A.; Maciejewski, W.; Atkinson, J.; Batcheldor, D.; Binney, J.; Carollo, M.; Dressel, L.; Ford, H.; et al.**, "Is There Really a Black Hole at the Center of NGC 4041? Constraints from Gas Kinematics", 2003, ApJ, 586, 868 (partly based on observations made with NICS@TNG)

**Masetti, N.; Palazzi, E.; Pian, E.; Simoncelli, A.; Hunt, L. K.; Maiorano, E.; Levan, A.; Christensen, L.; Rol, E.; Savaglio, S.; et al.**, "Optical and near-infrared observations of the GRB020405 afterglow", 2003, A&A, 404, 465 (partly based on observations made with DOLoRes@TNG)

**Pian, E.**, "Optical Observations of gamma-Ray Burst Afterglows", 2003, Supernovae and Gamma-Ray Bursters. Edited by K. Weiler, Lecture Notes in Physics, vol. 598,

p.343-366 (partly based on observations made with DOLoRes@TNG)

**Randich, S.; Sestito, P.; Pallavicini, R.**, "Evolution of lithium beyond the solar age: a Li survey of the old open cluster NGC 188", 2003, A&A, 399, 133 (based on observations made with SARG@TNG)

**Rigon, L.; Turatto, M.; Benetti, S.; Pastorello, A.; Cappellaro, E.; Aretxaga, I.; Vega, O.; Chavushyan, V.; Patat, F.; Danziger, I. J.; Salvo, M.**, "SN 1999E: another piece in the supernova-gamma-ray burst connection puzzle", 2003, MNRAS, 340, 191 (partly based on observations made with DOLoRes@TNG)

**Santos, N. C.; Israelian, G.; Mayor, M.; Rebolo, R.; Udry, S.**, "Statistical properties of exoplanets. II. Metallicity, orbital parameters, and space velocities", 2003, A&A, 398, 363 (partly based on observations made with SARG@TNG)

**Saracco, P.; Longhetti, M.; Severgnini, P.; Della Ceca, R.; Mannucci, F.; Bender, R.; Drory, N.; Feulner, G.; Ghinassi, F.; Hopp, U.; Maraston, C.**, "Massive z~1.3 evolved galaxies revealed", 2003, A&A, 398, 127 (based on observations made with NICS@TNG)

**Sazhin, M.; Longo, G.; Capaccioli, M.; Alcalá, J. M.; Silvotti, R.; Covone, G.; Khovanskaya, O.; Pavlov, M.; Pannella, M.; Radovich, M.; Testa, V.**, "CSL-1: chance projection effect or serendipitous discovery of a gravitational lens induced by a cosmic string?", 2003, MNRAS, 343, 353 (partly based on observations made with DOLoRes@TNG)

**Severgnini, P.; Caccianiga, A.; Braito, V.; Della Ceca, R.; Maccacaro, T.; Wolter, A.; Sekiguchi, K.; Sasaki, T.; Yoshida, M.; Akiyama, M.; et al.**, "XMM-Newton observations reveal AGN in apparently normal galaxies", 2003, A&A, 406, 483 (partly based on observations made with DOLoRes@TNG)

**Shepherd, D. S.; Testi, L.; Stark, D.**, "Clustered star formation in W75N", 2003, ApJ, 584, 882 (partly based on observations made with NICS@TNG)

**Valentini, G.; Di Carlo, E.; Massi, F.; et al.**, "Optical and near-infrared photometry of the Type Ia Supernova 2000E in NGC 6951", 2003, ApJ, 595, 779 (partly based on observations made with ARNICA@TNG)

# Publications 2003 on international non-refereed journals

**Antonelli, L. A.; Testa, V.; Fugazza, D.; Pedani, M.; Piro, L.; Covino, S.; Masetti, N.; Malesani, D.**, "GRB031220(=H2976): TNG r-band observation of the Chandra X-ray sources", 2003, GRB Coordinates Network, Circular Service, 2503, 1

**Barbieri, C.**, "Planets, Moons and Minor Bodies of the Solar System", 2003, Mem. SAIt Suppl., 3, 12

**Beers, T. C.; Lucatello, S.; Gratton, R. G.; Carretta, E.; Christlieb, N.; Cohen, J. G.**, "The Mystery of the Frequency of Ch-Stars at Low [FE/H]", 2003, *Elemental Abundances in Old Stars and Damped Lyman-α Systems*, 25th meeting of the IAU, Joint Discussion 15, 22 July 2003, Sydney, Australia

**Bonoli, C.; Giro, E.; Conconi, P.; Zitelli, V.**, "Tunable filter for the TNG", 2003, *Instrument Design and Performance for Optical/Infrared Ground-based Telescopes*. Edited by M. Iye and A. F. M. Moorwood. Proceedings of the SPIE, Volume 4841, pp. 494-500

**Campins, H.; Licandro, J.; Guerra, J. C.; Chamberlain, M.; Pantin, E.**, "Variations in the Nuclear Spectra of Comet 28P/Neujmin 1", 2003, American Astronomical Society, DPS meeting #35, #47.02

**Caproni, A.; Pucillo, M.; Smareglia, R.; Zacchei, A.; Oliva, E.**, "WSSL: the Workstation Software System under Linux", 2003, Mem SAIt Suppl, 3, 355

**Carollo, D.; Spagna, A.; Lattanzi, M. G.; Smart, R. L.; Hodgkin, S. T.; McLean, B. J.**, "Faint and peculiar objects in GAIA: results from GSC-II", 2003, *GAIA Spectroscopy: Science and Technology*, ASP Conference Proceedings, Vol. 298, held 9-12 September 2002 at La Residenza del Sole Congress Center, Gressoney St. Jean, Aosta, Italy. Edited by U. Munari. ISBN: 1-58381-145-1, p. 375

**D'Onghia, E.; Marchesini, D.; Chincarini, G.; Firmani, C.; Molinari, E.; Conconi, P.; Zacchei, A.**, "Dark Matter in the Center of LSB and Dwarf Galaxies from Hα Observations", 2003, *The Mass of Galaxies at Low and High Redshift*. Proceedings of the ESO Workshop held in Venice, Italy, 24-26 October 2001, p. 51

**de Leon Cruz, J.; Licandro, J.; Serra-Ricart, M.**, "Visible and Near Infrared Spectroscopic and Spectrophotometric Observations of NEOs", 2003, American Astronomical Society, DPS meeting #35, #34.07

**Della Ceca, R.; Maccacaro, T.; Fugazza, D.; Pedani, M.; Cecconi, M.; Fiore, F.; Antonelli, L. A.; Covino, S.; Pian, E.; Masetti, N.**, "GRB030329: optical spectroscopy with the TNG", 2003, GRB Coordinates Network, Circular Service, 2015, 1

**Diolaiti, E.; Tozzi, A.; Ragazzoni, R.; Ferruzzi, D.; Vernet-Viard, E.; Esposito, S.; Farinato, J.; Ghedina, A.; Riccardi, A.**, "Some novel concepts in multipyramid wavefront sensing", 2003, *Adaptive Optical System Technologies II*. Edited by P. L. Wizinowich and D. Bonaccini. Proceedings of the SPIE, Volume 4839, pp. 299-306

**Ecuvillon, A.; Gonzalez, J.; Israelian, G.; Santos, N.; Mayor, M.; Garcia Lopez, R.; Rebolo, R.; Randich, S.**, "Abundance of Volatiles N, O and S", 2003, *Stars as Suns: Activity, Evolution and Planets*, International Astronomical Union. Symposium no. 219, held 21-25 July, 2003 in Sydney, Australia

**Farinato, J.; Ragazzoni, R.; Diolaiti, E.; Vernet-Viard, E.; Baruffolo, A.; Arcidiacono, C.; Ghedina, A.; Cecconi, M.; et al.**, "Layer oriented adaptive optics: from drawings to metal", 2003, *Adaptive Optical System Technologies II*. Edited by P. L. Wizinowich and D. Bonaccini. Proceedings of the SPIE, Volume 4839, pp. 588-599

**Fugazza, D.; Antonelli, L. A.; Fiore, F.; Covino, S.; Ghisellini, G.; Pian, E.; Masetti, N.; Buzzoni, A.; Tessicini, G.**, "GRB030328: optical photometry at the TNG", 2003, GRB Coordinates Network, Circular Service, 1982, 1

**Fugazza, D.; Fiore, F.; Cocchia, M.; Antonelli, L. A.; Covino, S.; Pian, E.; Masetti, N.; Buzzoni, A.; Tessicini, G.**, "GRB030328: optical spectroscopy at TNG", 2003, GRB Coordinates Network, Circular Service, 1983, 1

**Ghedina, A.; Cecconi, M.; Ragazzoni, R.; Farinato, J.; et al.**, "On Sky Test of the Pyramid Wavefront Sensor", 2003, *Adaptive Optical System Technologies II*. Edited by P. L. Wizinowich and D. Bonaccini. Proceedings of the SPIE, Volume 4839, pp. 869-877

**Ghigo, M.; Diolaiti, E.; Perennes, F.; Ragazzoni, R.**, "Use of the LIGA process for the production of pyramid wavefront sensors for adaptive optics in astronomy", 2003, *Astronomical Adaptive Optics Systems and Applications*. Edited by R. K. Tyson, and M. Lloyd-Hart. Proceedings of the SPIE, Volume 5169, pp. 55-61

**Giro, E.; Bonoli, C.; Leone, F.; Molinari, E.; Pernechele, C.; Zacchei, A.**, " Polarization properties at the Nasmyth focus of the alt-azimuth TNG telescope", 2003, *Polarimetry in Astronomy*. Edited by S. Fineschi. Proceedings of the SPIE, Volume

4843, pp. 456-464

**Gratton, R.; Carretta, E.; Claudi, R. U.; Desidera, S.; Lucatello, S.; Bonanno, G.; Cosentino, R.; Scuderi, S., et al.**, "The SARG Planet Search: Hunting for Planets Around Stars in Wide Binaries", 2003, *Scientific Frontiers in Research on Extrasolar Planets*, ASP Conference Series, Vol 294, Edited by D. Deming and S. Seager (San Francisco: ASP) ISBN: 1-58381-141-9, pp. 47-50

**Lazzaro, D.; Duffard, R.; Licandro, J.; De Sanctis, M. C.; Capria, M. T.**, "Mineralogical Analysis of Basaltic Asteroids in the Neighbourhood of (4) Vesta", 2003, American Astronomical Society, DPS meeting #35, #34.13

**Leone, F.**, "The High Resolution Spectropolarimeter of the Italian Telescopio Nazionale Galileo", 2003, *Solar Polarization*. Proceedings of the Conference held 30 September - 4 October, 2002 at Tenerife, Canary Islands, Spain. Edited by J. Trujillo-Bueno and J. Sanchez Almeida. ASP Conference Proceedings, Vol. 307. San Francisco: Astronomical Society of the Pacific, p. 51

**Leone, F.; Bruno, P.; Cali, A.; Claudi, R.; Cosentino, R.; Gentile, G.; Gratton, R.; Scuderi, S.**, "High-resolution spectropolarimetry at the Telescopio Nazionale Galileo", 2003, *Polarimetry in Astronomy*. Edited by S. Fineschi. Proceedings of the SPIE, Volume 4843, pp. 465-475

**Licandro, J.; Campins, H.; de Leon Cruz, J.; Gil-Hutton, R.; Lara-Lopez, L. M.**, "Near infrared spectroscopy of TNOs, Centaurs and comet nuclei", 2003, American Astronomical Society, DPS meeting #35, #39.11

**López García, Z.; Malaroda, S.; González, E. P.; Tapia Vega, R.; Leone, F.**, "Abundancias químicas, campo magnético y composición isotópica en estrellas CP del grupo HgMn. I. HD 144206", 2003, Boletín de la Asociación Argentina de Astronomía, vol.46, p. 41

**Magazzù, A.; Dougados, C.; Licandro, J.; Martín, E. L.; Magnier, E. A.; Ménard, F.**, "Infrared Spectra of Brown Dwarf Candidates in Taurus", 2003, *Brown Dwarfs*, Proceedings of IAU Symposium #211, held 20-24 May 2002 at University of Hawaii, Honolulu, Hawaii. Edited by E. Martin. San Francisco: Astronomical Society of the Pacific, 2003, p. 7

**Mazzoleni, R.; Zerbi, F. M.; Held, E. V.; Ciroi, S.; Conconi, P.; et al.**, "Retrofitting focal reducer spectrographs with removable integral field units", 2003, *Specialized Optical Developments in Astronomy*. Edited by E. Atad-Ettedgui, and S. D'Odorico. Proceedings of the SPIE, Volume 4842, pp. 219-230

**Oliva, E.**, "NICS, the Near Infrared Camera-Spectrometer of the TNG", 2003, Mem. SAIt, 74, 118

**Pallavicini, R.**, "Galactic open clusters: key tracers of stellar structure and evolution", 2003, Mem. SAIt Suppl., 3, 74

**Pasian, F.; Smareglia, R.; Vuerli, C.; Zacchei, A.; Lama, N.; Benacchio, L.**, "INAF Archives in the Framework of the Astronomical Data Grid", 2003, Mem. SAIt Suppl., 3, 368

**Ranalli, P.; Origlia, L.; Maiolino, R.; Comastri, A.**, "α-Elements Enhancement in Star Bursts: the Case of M82", 2003, *Atomic Data for X-Ray Astronomy*, 25th meeting of the IAU, Joint Discussion 17, 22 July 2003, Sydney, Australia

**Ruiz-Lapuente, P.; Balastegui, A.; Sainton, G.; Pascoal, R.; Amanullah, R.**, "Supernovae 2002li-2002ls", 2003, IAU Circular, 8181, 1 (partly based on observations made with DOLoRes@TNG)

**Saracco, P.; Longhetti, M.; Severgnini, P.; Della Ceca, R.; Mannucci, F.; Ghinassi, F.; Drory, N.; Feulner, G.; Bender, R.; Maraston, C.; Hopp, U.**, "TESIS - The TNG EROs Spectroscopic Identification Survey", 2003, *Galaxy Evolution: Theory & Observations* (Eds. V. Avila-Reese, C. Firmani, C. S. Frenk & C. Allen), Revista Mexicana de Astronomía y Astrofísica (Serie de Conferencias) Vol. 17, pp. 249-249

**Smareglia, R.; Becciani, U.; Caproni, A.; Gheller, C.; Guerra, J. C.; Lama, N.; Longo, G.; Pasian, F.; Zacchei, A.**, "The pilot project for the TNG Long-Term Archive", 2003, Mem. SAIt, 74, 514

**Testa, V.; Cocozza, G.; Melandri, A.; Antonelli, L. A.; Malesani, D.; Ghinassi, F.; Fugazza, D.; di Tomaso, S.; Covino, S.; Masetti, N.; Pian, E.**, "GRB030329: VR photometry at TNG", 2003, GRB Coordinates Network, Circular Service, 2141, 1

**Testa, V.; Fugazza, D.; Della Valle, M.; Malesani, D.; Mason, E.; Pian, E.; et al.**, "GRB 021211: R-band observations at late-time", 2003, GRB Coordinates Network, Circular Service, 1821, 1

**Testi, L.; Natta, A.; Baffa, C.; Comoretto, G.; Gennari, S.; Ghinassi, F.; Licandro, J.; Magazzù, A.; Oliva, E.; D'Antona, F.**, "An Efficient Low-Resolution

NIR Classification Scheme for M, L, and T dwarfs and Its Application to Young Bds", 2003, *Brown Dwarfs*, Proceedings of IAU Symposium #211, held 20-24 May 2002 at University of Hawaii, Honolulu, Hawaii. Edited by E. Martín. San Francisco: Astronomical Society of the Pacific, p. 359

**Testi, L.; Natta, A.; Comerón, F.; Oliva, E.; D'Antona, F.**, "Disks in Brown Dwarf Systems", 2003, *Brown Dwarfs*, Proceedings of IAU Symposium #211, held 20-24 May 2002 at University of Hawaii, Honolulu, Hawaii. Edited by E. Martín. San Francisco: Astronomical Society of the Pacific, p. 127

**Turatto, M.**, "Local supernovae", 2003, Mem. SAIt Suppl., v.3, p. 109

**Viotti, R.F.; Claudi, R.; Rossi, C.; Iijima, T.**, "High Resolution Observations of Emission Line Stars", 2003, *Frontiers of High Resolution Spectroscopy*, 25th meeting of the IAU, Joint Discussion 20, 23-24 July 2003, Sydney, Australia

**Vitali, F.; Cianci, E.; Foglietti, V.; Lorenzetti, D.**, "Toward the fabrication of silicon grisms for high resolution spectroscopy in the near infrared", 2003, Mem SAIt, 74, 197

# Publications 2002 on international journals with "referee"

**Bellazzini, M.; Ferraro, F.; Ibata, R.**, "The stellar population of NGC5634. A globular cluster in the Sagittarius dSph Stream?", 2002, AJ, 124, 915 (based on observations made with DOLoRes@TNG)

**Bellazzini, M.; Ferraro, F. R.; Origlia, L.; Pancino, E.; Monaco, L.; Oliva, E.**, "The Draco and Ursa Minor Dwarf Spheroidal Galaxies: A Comparative Study", 2002, AJ, 124, 3222 (based on observations made with DOLoRes@TNG)

**Capetti, A.; Zamfir, S.; Rossi, P.; Bodo, G.; Zanni, C.; Massaglia, S.**, "On the origin of X-shaped radio-sources: New insights from the properties of their host galaxies", 2002, A&A, 394, 39 (based on observations made with NICS@TNG)

**Carollo, D.; Hodgkin, S. T.; Spagna, A.; Smart, R. L.; Lattanzi, M. G.; McLean, B. J.; Pinfield, D. J.**, "Discovery of a peculiar DQ white dwarf", 2002, A&A, 393, L45 (partly based on observations made with NICS@TNG)

**Conconi, P.; Molinari, E.; Crimi, G.**, "Gravitational eccentric correction optics (GECO): an optical-gravitational device to compensate for flexures in astronomical spectrographs", 2002, Applied Optics, Volume 41, Issue 1, pp. 193-196 (based on data taken with DOLoRes@TNG)

**Corsini, E. M.; Pizzella, A.; Bertola, F.**, "The orthogonal gaseous kinematical decoupling in the Sa spiral NGC 2855", 2002, A&A, 382, 488 (partly based on observations made with OIG@TNG)

**Dallacasa, D.; Falomo, R.; Stanghellini, C.**, "Optical identifications of High Frequency Peakers", 2002, A&A, 382, 53 (based on observations made with DOLoRes@TNG)

**Debattista, V. P.; Corsini, E. M.; Aguerri, J. A. L.**, "A fast bar in the post-interaction galaxy NGC 1023", 2002, MNRAS, 332, 65 (based on observations made with DOLoRes@TNG)

**Di Carlo, E.; Massi, F.; Valentini, G.; Di Paola, A.; D'Alessio, F.; Brocato, E.; Guidubaldi, D.; Dolci, M.; Pedichini, F.; Speziali, R.; et al.**, "Optical and infrared observations of the supernova SN 1999el", 2002, ApJ, 573, 144 (partly based on observations made with OIG@TNG)

**Ghinassi, F.; Licandro, J.; Oliva, E.; Baffa, C.; Checcucci, A.; Comoretto, G.; Gennari, S.; Marcucci, G.**, "Transmission curves and effective refraction indices of MKO near infrared consortium filters at cryogenic temperatures", 2002, A&A, 386, 1157 (based on data taken with NICS@TNG)

**Grazian, A.; Omizzolo, A.; Corbally, C.; Cristiani, S.; Haehnelt, M. G.; Vanzella, E.**, "The Asiago-ESO/RASS QSO Survey. II. The Southern Sample", 2002, AJ, 124, 2955 (partly based on observations made with DOLoRes@TNG)

**Israel, G. L.; Hummel, W.; Covino, S.; Campana, S.; Appenzeller, I.; Gassler, W.; Mantel, K.-H.; Marconi, G.; Mauche, C. W.; Munari, U.; et al.**, "RX J0806.3+1527: A double degenerate binary with the shortest known orbital period (321s)", 2002, A&A, 383, L13 (partly based on observations made with DOLoRes@TNG)

**Licandro, J.; Ghinassi, F.; Testi, L.**, "Infrared spectroscopy of the largest known trans- neptunian object 2001 KX$_{76}$", 2002, A&A, 388, L9 (based on observations made with NICS@TNG)

**Licandro, J.; Guerra, J.C.; Campins, H., et al.**, "The Surface of Cometary Nulcei Related Minor Icy Bodies", 2002, EM&P, 90, 495 (based on observations made with NICS@TNG)

**Maiolino, R.; Vanzi, L.; Mannucci, F.; Cresci, G.; Ghinassi, F.; Della Valle, M.**, "Discovery of two infrared supernovae: a new window on the SN search", 2002, A&A, 389, 84 (partly based on observations made with ARNICA and NICS@TNG)

**Mannucci, F.; Pozzetti, L.; Thompson, D.; Oliva, E.; Baffa, C.; Comoretto, G.; Gennari, S.; Lisi, F.**, "The relative abundances of ellipticals and starbursts among the Extremely Red Galaxies", 2002, MNRAS, 329, L57 (based on observations made with NICS@TNG)

**Marchesini, D.; D'Onghia, E.; Chincarini, G.; Firmani, C.; Conconi, P.; Molinari, E.; Zacchei, A.**, "Hα Rotation Curves: The Soft Core Question", 2002, ApJ, 575, 801 (based on observations made with DOLoRes@TNG)

**Nagar, N. M.; Oliva, E.; Marconi, A.; Maiolino, R.**, "NGC 5506 unmasked as a Narrow Line Seyfert 1:. A direct view of the broad line region using near- IR spectroscopy", 2002, A&A, 391, 21 (based on observations made with NICS@TNG)

**Natta, A.; Testi, L.; Comeron, F.; Oliva, E.; D'Antona, F.; Baffa, C.; Comoretto,**

**G.; Gennari, S.**, "Exploring Brown Dwarf Disks in rho-Ophiuchi", 2002, A&A, 393, 597 (partly based on observations made with NICS@TNG)

**Peixinho, N.; Doressoundiram, A.; Romon-Martin, J.**, "Visible-IR colors and lightcurve analysis of two bright TNOs: 1999 $TC_{36}$ and 1998 $SN_{165}$", 2002, New Astronomy, Volume 7, Issue 6, p. 359-367 (partly based on observations made with ARNICA@TNG)

**Testi, L.; Natta, A.; Oliva, E.; D'Antona, F.; Comeron, F.; Baffa, C.; Comoretto, G.; Gennari, S.**, "A young very low-mass object surrounded by warm dust", 2002, ApJ, 571, L155 (partly based on observations made with NICS@TNG)

**Tozzi, G. P.; Licandro, J.**, "Visible and infrared images of C/1999 S4 (LINEAR) during the disruption of its nucleus", 2002, Icarus, 157, 187 (based on observations made with NICS@TNG)

## Publications <span style="color:red">2002</span> on international non-refereed journals

**Brunetti, G.; Comastri, A.; Dallacasa, D.; Bondi, M.; Pedani, M.; Setti, G.**, "Chandra detects inverse Compton emission from the radio galaxy 3C 219", 2002, Proceedings of the Symposium *New Visions of the X-ray Universe in the XMM-Newton and Chandra Era*, ESTEC 2001

**Capria, M. T.; Cremonese, G.; Boattini, A.; de Sanctis, M. C.; D'Abramo, G.; Buzzoni, A.**, "High resolution spectroscopy of comet C/2002 C1 Ikeya-Zhang with SARG at TNG", 2002, Proceedings of *Asteroids, Comets, Meteors* - ACM 2002. International Conference, 29 July - 2 August 2002, Berlin, Germany. Ed. B. Warmbein. ESA SP-500. Noordwijk, Netherlands: ESA Publications Division, ISBN 92-9092-810-7, p. 693 - 696

**Castro Ceron, J. M.; Pedani, M.; Gorosabel, J.; Castro-Tirado, A.J.**, "GRB 020603, R-band observation", 2002, GRB Coordinates Network, Circular Service, 1413, 1

**Cellino, A.; Diolaiti, E.; Ghedina, A.; Hestroffer, D.; Ragazzoni, R.; Tanga, P.**, "Speckle interferometry observations of main belt asteroids at TNG", 2002, Proceedings of *Asteroids, Comets, Meteors* - ACM 2002. International Conference, 29 July - 2 August 2002, Berlin, Germany. Ed. B. Warmbein. ESA SP-500. Noordwijk, Netherlands: ESA Publications Division, ISBN 92-9092-810-7, 2002, p. 497 – 500

**Conconi, P.; Molinari, E.; Zerbi, F. M.; et al.**, "VPHG-based upgrade of the low-resolution spectrograph DOLoRes at the Italian Galileo National Telescope", 2002, Proc. SPIE Vol. 4485, p. 445-452, *Optical Spectroscopic Techniques, Remote Sensing, and Instrumentation for Atmospheric and Space Research IV*, A. M. Larar; M. G. Mlynczak; Eds.

**Covino, S.; Malesani, D.; Ghisellini, G.; Stefanon, M.; Tavecchio, F.; Covone, G.; et al.**, "XRF020903: TNG and asiago photometry", 2002, GRB Coordinates Network, Circular Service, 1563, 1

**Cremonese, G.; Boattini, A.; Capria, M. T.; De Sanctis, M. C.; D'Abramo, G.; Buzzoni, A.**, "Comet C/2002 C1 (Ikeya-Zhang)", 2002, IAU Circular, 7914, 2

**Cresci, G.; Mannucci, F.; Maiolino, R.; Della Valle, M.; Ghinassi, F.**, "Supernova 1999gw in UGC 4881", 2002, IAU Circular, 7784, 1 (discovery of an apparent supernova with ARNICA@TNG)

**Danziger, J.; Della Valle, M.; Palazzi, E.; Pian, E.**, "Supernova 2002ap in M74", 2002, IAU Circular, 7838, 2

**Dotto, E.; Leyrat, C.; Romon, J.; Barucci, M. A.; Licandro, J.; de Bergh, C.**, " A portrait of the Centaur 10199 Chariklo", 2002, American Astronomical Society, DPS Meeting #34, #07.02; Bulletin of the American Astronomical Society, Vol. 34, p. 845

**Fornasier, S.; Barucci, M. A.; Binzel, R. P.; Birlan, M.; Fulchignoni, M.; et al.**, "Spectrophotometric observations of 4979 Otawara, target of the Rosetta space mission", 2002, American Astronomical Society, DPS Meeting #34, #14.11; Bulletin of the American Astronomical Society, Vol. 34, p. 860

**Gratton, R.; Bonanno, G.; Brocato, E.; Carretta, E.; Claudi, R.; Cosentino, R.; et al.**, "SARG Extra Solar Planet Search", 2002, Proceedings of the First Eddington Workshop on *Stellar Structure and Habitable Planet Finding*, 11 - 15 June 2001, Córdoba, Spain. Editor: B. Battrick, Scientific editors: F. Favata, I. W. Roxburgh & D. Galadi. ESA SP-485, Noordwijk: ESA Publications Division, ISBN 92-9092-781-X, p. 265 – 268

**Israel, G. L.; Stella, L.; Hummel, W.; Covino, S.; Campana, S.**, "RX J0806.3+1527", 2002, IAU Circular, 7835, 2

**Knop, R.**, "Supernovae 2002gi, 2002gj, 2002gk, 2002gl, 2002gm", 2002, IAU

Circular, 7993, 1

**Longo, G.; Donalek, C.; Raiconi, G.; Staiano, A.; Tagliaferri, R.; Sessa, S.; Pasian, F.; Smareglia, R.; Volpicelli, A.**, "Data mining of large astronomical databases with neural tools", 2002, *Astronomical Data Analysis II*. Edited by Starck, J.-L.; Murtagh, F. D. Proceedings of the SPIE, Volume 4847, pp. 265-276

**Malesani, D.; Covino, S.; Fugazza, D.; Barrena, R.; Pian, E.; Masetti, N.**, "GRB020813: V-band decay slope", 2002, GRB Coordinates Network, Circular Service, 1500, 1

**Malesani, D.; Covino, S.; Ghisellini, G.; Lazzati, D.; Cecconi, M.; Fugazza, D.; Guerra, J. C.; Masetti, N.; Pian, E.**, "GRB021004: optical observations and predicted break time", 2002, GRB Coordinates Network, Circular Service, 1607, 1

**Malesani, D.; Stefanon, M.; Covino, S.; Ghisellini, G.; Lazzati, D.; Rossi, E.; Fugazza, D.; Guerra, J. C.; Pedani, M.; Masetti, N.; Pian, E.**, "GRB 021004: VRI observations at TNG: possible break", 2002, GRB Coordinates Network, Circular Service, 1645, 1

**Marchesini, D.; D'Onghia, E.; Chincarini, G.; Firmani, C.; Conconi, P.; Molinari, E.; Zacchei, A.**, "Soft Cores in Late-Type Dwarf and LSB Galaxies from Hα Observations", 2002, *Tracing Cosmic Evolution with Galaxy Clusters*. ASP Conference Proceedings, Vol. 268. Edited by S. Borgani, M. Mezzetti, and R. Valdarnini. ISBN: 1-58381-108-7 San Francisco, Astronomical Society of the Pacific, p. 407

**Palazzi, E.; Masetti, N.; Pian, E.; Licandro, J.; Maiolino, R.; Saracco, P.; Fiore, F.**, "V band observations of GRB020405", 2002, GRB Coordinates Network, Circular Service, 1328, 1

**Porceddu, I.; Zitelli, V.; Buffa, F.; Ghedina, A.**, "Dust pollution monitoring at the TNG telescope", 2002, *Observatory Operations to Optimize Scientific Return III*. Edited by P. J. Quinn. Proceedings of the SPIE, Volume 4844, pp. 358-365

**Ragazzoni, R.; Esposito, S.; Ghedina, A.; Baruffolo, A.; Cecconi, M.; et al.**, "Pyramid wavefront sensor aboard AdOpt@TNG and beyond: a status report", 2002, Proc. SPIE Vol. 4494, p. 181-187, *Adaptive Optics Systems and Technology II*, R.K. Tyson; D. Bonaccini; M. C. Roggemann; Eds.

**Riello, M.; Benetti, S.; Altavilla, G.; Pastorello, A.; Turatto, M.; Cappellaro, E.;**

**Ruiz-Lapuente, P.; Matheson, T.; Calkins, M.; Chornock, R.**, "Supernovae 2002cj, 2002ck, 2002cp, 2002cs", 2002, IAU Circular, 7894, 4

**Smareglia, R.; Pasian, F.; Zacchei, A.; Caproni, A.; Vuerli, C.; Longo, G.; Becciani, U.; Gheller, C.**, "Archive systems for the TNG telescope: lessons learned in the VO perspective", 2002, Virtual Observatories. Edited by Szalay, Alexander S. Proceedings of the SPIE, Volume 4846, pp. 158-169

**Stefanon, M.; Covino, S.; Malesani, D.; Ghisellini, G.; Fugazza, D.; Tessicini, G.; Masetti, N.; Pian, E.,** "GRB 021004: Optical/NIR Observations", 2002, GRB Coordinates Network, Circular Service, 1623, 1

**Testi, L.; D'Antona, F.; Ghinassi, F.; Licandro, J.; Magazzù, A.; Natta, A.; Oliva, E.**, "NIR Low-Resolution Spectroscopy of L-Dwarfs: An Efficient Classification Scheme for Faint Dwarfs", 2002, *The Origins of Stars and Planets: The VLT View.* Proceedings of the ESO Workshop held in Garching, Germany, 24-27 April 2001, p. 187.

# Publications 2001 on international journals with "referee"

**Baffa, C.; Comoretto, G.; Gennari, S.; Lisi, F.; Oliva, E.; Biliotti, V.; Checcucci, A.; Gavrioussev, V.; Giani, E.; Ghinassi, F.; et al.**, "NICS: The TNG Near Infrared Camera Spectrometer", 2001, A&A, 378, 722 (based on data taken with NICS@TNG)

**Bortoletto, F.**, "The commissioning of the TNG telescope", 2001, New Astronomy Reviews, Volume 45, Issue 1-2, p. 37-40

**Bragaglia, A.; Carretta, E.; Gratton, R. G.; Tosi, M.; Bonanno, G.; Bruno, P.; Calì, A.; Claudi, R.; Cosentino, R.; Desidera, S.; et al.**, "Metal Abundances of Red Clumps Stars in Open Clusters: I. NGC 6819", 2001, AJ, 121, 327 (based on observations made with SARG@TNG)

**Brunetti, G.; Bondi, M.; Comastri, A.; Pedani, M.; Varano, S.; Setti, G.; Hardcastle, M. J.**, "Chandra detection of the radio and optical double Hot Spot of 3C 351", 2001, ApJ, 561, L157 (partly based on observations made with DOLoRes@TNG)

**Castro-Tirado, A. J.; Sokolov, V. V.; Gorosabel, J.; Castro Ceron, J. M.; Greiner, J.; Wijers, R. A. M. J.; Jensen, B. L.; Hjorth, J.; Toft, S.; Pedersen, H.; et al.**, "The extraordinarily bright optical afterglow of GRB 991208 and its host galaxy", 2001, A&A, 370, 398 (partly based on observations made with OIG@TNG)

**Doressoundiram, A.; Barucci, M. A.; Romon, J.; Veillet, C.**, "Multicolor Photometry of Trans-neptunian Objects", 2001, Icarus, 154, 277 (partly based on observations made with OIG@TNG)

**Fynbo, J. U.; Jensen, B. L.; Gorosabel, J.; Hjorth, J.; Pedersen, H.; Moller, P.; Abbott, T.; Castro-Tirado, A. J.; Delgado, D.; Greiner, J.; et al.**, "Detection of the optical afterglow of GRB 000630: Implications for dark bursts", 2001, A&A, 369, 373 (partly based on observations made with OIG@TNG)

**Gavazzi, G.; Zibetti, S.; Boselli, A.; Franzetti, P.; Scodeggio, M.; Martocchi, S.**, "1.65 micron (H-band) surface photometry of galaxies. VII. Dwarf galaxies in the Virgo Cluster", 2001, A&A, 372, 29 (partly based on observations made with ARNICA@TNG)

**Gil-Hutton, R.; Licandro, J.**, "VR Photometry of Sixteen Kuiper Belt Objects", 2001, Icarus, 152, 246 (partly based on observations made with OIG@TNG)

**Gratton, R. G.; Bonanno, G.; Bruno, P.; Cali, A.; Claudi, R. U.; Cosentino, R.; Desidera, S.; et al.**, "SARG: the high resolution spectrograph of TNG", 2001, Experimental Astronomy, 12, 107

**Gratton, R. G.; Bonanno, G.; Claudi, R. U.; Cosentino, R.; Desidera, S.; Lucatello, S.; Scuderi, S.**, "Non-interacting main-sequence binaries with different chemical compositions: Evidences of infall of rocky material?", 2001, A&A, 377, 123 (based on observations made with SARG@TNG)

**Gutierrez, P. J.; Ortiz, J. L.; Alexandrino, E.; Roos-Serote, M.; Doressoundiram, A.**, "Short term variability of Centaur 1999UG5", 2001, A&A, 371, L1 (partly based on observations made with DOLoRes@TNG)

**Licandro, J.; Oliva, E.; Di Martino, M.**, "NICS-TNG infrared spectroscopy of trans-neptunian objects 2000 EB173 and 2000 WR106", 2001, A&A, 373, L29 (based on observations made with NICS@TNG)

**Maiolino, R.; Mannucci, F.; Baffa, C.; Gennari, S.; Oliva, E.**, "Discovery of strong CIV absorption in the highest redshift quasar", 2001, A&A, 372, L5 (based on observations made with NICS@TNG)

**Marchetti, E.; Faraggiana, R.; Bonifacio, P.**, "A speckle interferometry survey of λ Bootis stars", 2001, A&A, 370, 524 (based on observations made with the SPECKLE CAMERA@TNG)

**Masetti, N.; Palazzi, E.; Pian, E.; Mannucci, F.; Antonelli, L. A.; Di Paola, A.; Saracco, P.; Savaglio, S.; Amati, L.; Bartolini, C.; et al.**, "GRB010222: afterglow emission from a rapidly decelerating shock", 2001, A&A, 374, 382 (partly based on observations made with DOLoRes@TNG)

**Munari, U.; Tomov, T.; Yudin, B. F.; Marrese, P. M.; Zwitter, T.; Gratton, R. G.; Bonanno, G.; Bruno, P.; Calì, A.; Claudi, R. U.; et al.**, "Discovery of a bipolar and highly variable mass outflow from the symbiotic binary StHα 190", 2001, A&A, 369, L1 (partly based on observations made with SARG@TNG)

**Muñoz, J. A.; Mediavilla, E.; Falco, E. E.; Oscoz, A.; Barrena, R.; McLeod, B. A.; Abajas, C.; Alcalde, D.; Serra-Ricart, M.; Motta, V.**, "Subarcsecond Optical Images of the Radio Gravitational Lens B1152+199", 2001, ApJ, 563, 107 (based on observations made with OIG@TNG)

**Oliva, E.; Marconi, A.; Maiolino, R.; Testi, L.; Mannucci, F.; Ghinassi, F.; Licando, J.; Origlia, L.; Baffa, C.; Checcucci, A.; et al.**, "NICS-TNG infrared spectroscopy of NGC1068: the first extragalactic measurement of [PII] and a new tool to constrain the origin of [FeII] line emission in galaxies", 2001, A&A, 369, L5 (based on observations made with NICS@TNG)

**Ragazzoni, R.; Cappellaro, E.; Benetti, S.; Turatto, M.; Sabbadin, F.**, "3-D ionization structure (in stereoscopic view) of planetary nebulae: the case of NGC 1501", 2001, A&A, 369, 1088 (partly based on observations made with SARG@TNG)

**Romon, J.; de Bergh, C.; Barucci, M. A.; Doressoundiram, A.; Cuby, J.-G.; Le Bras, A.; Doutè, S.; Schmitt, B.**, "Photometric and spectroscopic observations of Sycorax, satellite of Uranus", 2001, A&A, 376, 310 (partly based on observations made with OIG and ARNICA@TNG)

**Testi, L.; D'Antona, F.; Ghinassi, F.; Licandro, J.; Magazzù, A.; Maiolino, R.; Mannucci, F.; Marconi, A.; Nagar, N.; Natta, A.; Oliva, E.**, "NICS-TNG low-resolution 0.85-2.45 um spectra of L-Dwarfs: a near-infrared spectral classification scheme for faint dwarfs", 2001, ApJ, 552, L147 (based on observations made with NICS@TNG)

# Publications 2001 on international non-refereed journals

**Antonelli, L. A.; Mannucci, F.; Pian, E.; Testa, V.; di Paola, A.; Stella, L.; Burud, I.; Fruchter, A.; Rhoads, J.; Masetti, N.; et al.**, "GRB 011030: TNG K-band observation", 2001, GRB Coordinates Network, Circular Service, 1146, 1

**Baffa, C.; Gennari, S.; Lisi, F.; et al.**, "NICS, the IR imager and spectrograph", 2001, Conference Proceedings of *Scientific Dedication of the Telescopio Nazionale Galileo*, La Palma 3-5 November 2000, G. Setti, M. Rodonò, Eds.

**Barbieri, C.**, "The design and construction phases", 2001, Conference Proceedings of *Scientific Dedication of the Telescopio Nazionale Galileo*, La Palma 3-5 November 2000, G. Setti, M. Rodonò, Eds.

**Benetti, S.; Altavilla, G.; Cappellaro, E.; Pastorello, A.; Turatto, M.; Cosentino, R.; Zacchei, A.; Danziger, I. J.; Patat, F.**, "Supernova 2001bg in NGC 2608", 2001, IAU Circular, 7639, 2

**Bonanno, G.; Bonoli, C.; Bortoletto, F.; et al.**, "The CCD controllers and detectors", 2001, Conference Proceedings of *Scientific Dedication of the Telescopio Nazionale Galileo*, La Palma 3-5 November 2000, G. Setti, M. Rodonò, Eds.

**Bonoli, C.; Corcione, L.; Fantinel, D.; Gardiol, D.**, "The tracking and control systems", 2001, Conference Proceedings of *Scientific Dedication of the Telescopio Nazionale Galileo*, La Palma 3-5 November 2000, G. Setti, M. Rodonò, Eds.

**Bortoletto, F.; Benetti, S.; Bonanno, C.; et al.**, "The optical imager "Galileo" (OIG)", 2001, Conference Proceedings of *Scientific Dedication of the Telescopio Nazionale Galileo*, La Palma 3-5 November 2000, G. Setti, M. Rodonò, Eds.

**Bortoletto, F.; et al.**, "The comissioning phase", 2001, Conference Proceedings of *Scientific Dedication of the Telescopio Nazionale Galileo*, La Palma 3-5 November 2000, G. Setti, M. Rodonò, Eds.

**Campins, H.; Licandro, J.; Chamberlain, M.; Brown, R. H.**, "Constraints on the Surface Composition of Comet 28P/Neujmin 1", 2001, American Astronomical Society, DPS Meeting #33, #31.08; Bulletin of the American Astronomical Society, Vol. 33, p. 1094

**Catanzaro, G.; Leone, F.; Andre, M.; Sonnentrucker, P.**, "A spectroscopic study of the suspected chemically peculiar star HD207538. FUSE and SARG-TNG combined spectra", 2001, American Astronomical Society, 198th AAS Meeting, #45.03; Bulletin of the American Astronomical Society, Vol. 33, p. 846

**Cellino, A.; Diolaiti, E.; Ghedina, A.; Hestroffer, D.; Ragazzoni, R.; Tanga, P.**, "Asteroid Observations Using the Speckle Camera at TNG", 2001, American Astronomical Society, DPS Meeting #33, #61.02; Bulletin of the American Astronomical Society, Vol. 33, p. 1152

**Conconi, P.; Benetti, S.; Bergamini, U.; et al.**, "DOLoRes: the low-resolution optical imager and spectrograph", 2001, Conference Proceedings of *Scientific Dedication of the Telescopio Nazionale Galileo*, La Palma 3-5 November 2000, G. Setti, M. Rodonò, Eds.

**Fiorani, A.; Scaramella, R.; Lorenzetti, D.; Vitali, F.**, "NIR Visibility Function of Emission Lines with the Galileo OH Subtracted Spectrograph", 2001, *Mining the Sky*, Proceedings of the MPA/ESO/MPE Workshop held at Garching, Germany, 31 July-4 August, 2000. Edited by A. J. Banday, S. Zaroubi, and M. Bartelmann. Heidelberg: Springer-Verlag, p. 350

**Fiore, F.; Antonelli, L. A.; Ciliegi, P.; Comastri, A.; Giommi, P.; La Franca, F.; Maiolino, R.; Matt, G.; Molendi, S.; Perola, G. C.; Vignali, C.**, "The BeppoSAX hellas survey: On the nature of faint hard X-ray selected sources", 2001, *X-RAY ASTRONOMY: Stellar Endpoints,AGN, and the Diffuse X-ray Background*. Edited by N. E. White, G. Malaguti, and G. Palumbo. Melville, NY: American Institute of Physics, 2001. AIP Conference Proceedings, Volume 599, pp. 111-119

**Fiore, F.; Israel, G. L.; Antonelli, L. A.; Stella, L.; Covino, S.; Saracco, P.; Ghisellini, G.; Buzzoni, A.; Oliva, E.; Pian, E.; et al.**, "GRB 011211: TNG spectroscopic observations", 2001, GRB Coordinates Network, Circular Service, 1203, 1

**Fusi Pecci, F.; Zitelli, V.**, "The focal plane instruments' working groups", 2001, Conference Proceedings of *Scientific Dedication of the Telescopio Nazionale Galileo*, La Palma 3-5 November 2000, G. Setti, M. Rodonò, Eds.

**Giro, E.; Zacchei, A.**, "The user interfaces", 2001, Conference Proceedings of *Scientific Dedication of the Telescopio Nazionale Galileo*, La Palma 3-5 November 2000, G. Setti, M. Rodonò, Eds.

**Gratton, R.G.; Bonanno, G.; Bruno, P.; et al.**, "SARG: the high resolution spectrograph", 2001, Conference Proceedings of *Scientific Dedication of the Telescopio Nazionale Galileo*, La Palma 3-5 November 2000, G. Setti, M. Rodonò, Eds.

**Maccacaro, T.**, "User statistics and proposal selection", 2001, Conference Proceedings of *Scientific Dedication of the Telescopio Nazionale Galileo*, La Palma 3-5 November 2000, G. Setti, M. Rodonò, Eds.

**Mancini, D.; Brescia, M.**, "The differential image motion monitor", 2001, Conference Proceedings of *Scientific Dedication of the Telescopio Nazionale Galileo*, La Palma 3-5 November 2000, G. Setti, M. Rodonò, Eds.

**Mancini, D.; Schipani, P.**, "The drive and axes control system", 2001, Conference Proceedings of *Scientific Dedication of the Telescopio Nazionale Galileo*, La Palma 3-5 November 2000, G. Setti, M., Rodonò, Eds.

**Masetti, N.; Palazzi, E.; Pedani, M.; Magazzù, A.; Ghinassi, F.; Pian, E.**, "GRB010126, TNG r-band observations", 2001, GRB Coordinates Network, Circular Service, 926, 1

**Masetti, N.; Palazzi, E.; Pian, E.; Desidera, S.; Giro, E.; Frigo, A.; Falomo, R.; Zacchei, A.; Magazzù, A.; Ghinassi, F.; Pedani, M.,** "GRB010214, optical observations", 2001, GRB Coordinates Network, Circular Service, 954, 1

**Masetti, N.; Palazzi, E.; Pian, E.; Zacchei, A.; Magazzù, A.; Pedani, M.; Ghinassi, F.; Mignoli, M.,** "GRB010222, possible increase of optical decay slope", 2001, GRB Coordinates Network, Circular Service, 985, 1

**Oliva, E.,** "First scientific results", 2001, Conference Proceedings of *Scientific Dedication of the Telescopio Nazionale Galileo*, La Palma 3-5 November 2000, G. Setti, M. Rodonò, Eds.

**Ortolani, S.; Zitelli, V.,** "The weather and seeing monitoring", 2001, Conference Proceedings of *Scientific Dedication of the Telescopio Nazionale Galileo*, La Palma 3-5 November 2000, G. Setti, M. Rodonò, Eds.

**Pasian, F.; Smareglia, R.; Benacchio, L.,** "The data archives – at the telescope and beyond", 2001, Conference Proceedings of *Scientific Dedication of the Telescopio Nazionale Galileo*, La Palma 3-5 November 2000, G. Setti, M. Rodonò, Eds.

**Peixinho, N.; Doressoundiram, A.; Barucci, M. A.; Romon-Martin, J.,** "Portrait of two bright TNOs: 1999TC36 and 1998SN165", 2001, American Astronomical Society, DPS Meeting #33, #12.02; Bulletin of the American Astronomical Society, Vol. 33, p. 1046

**Porceddu, I.; Buffa, F.; Cocco, G.C.; Serrau, M.M.,** "An overview of the astroclimatic stations", 2001, Conference Proceedings of *Scientific Dedication of the Telescopio Nazionale Galileo*, La Palma 3-5 November 2000, G. Setti, M. Rodonò, Eds.

**Ragazzoni, R.,** "AdOpt@TNG: a success in progress", 2001, Conference Proceedings of *Scientific Dedication of the Telescopio Nazionale Galileo*, La Palma 3-5 November 2000, G. Setti, M. Rodonò, Eds.

**Rodonò, M.,** "From the kick-off to the operational phase", 2001, Conference Proceedings of *Scientific Dedication of the Telescopio Nazionale Galileo*, La Palma 3-5 November 2000, G. Setti, M. Rodonò, Eds.

**Rossi, C.; Viotti, R. F.; Gaeng, T.; Gratton, R.; Claudi, R.; Farisato, G.; Martorana, G.; et al.,** "The SARG very high resolution spectrum of P Cygni", 2001,

*P Cygni 2000: 400 Years of Progress*, ASP Conference Proceeding Vol. 233, Edited by M. de Groot and C. Sterken, San Francisco: Astronomical Society of the Pacific. ISBN: 1-58381-070-6, p. 109

**Sanchez-Lavega, A.; Acarreta, J. R.; Tanga, P.**, "Cloud structure in Saturn's South Polar Region", 2001, American Astronomical Society, DPS Meeting #33, #11.18; Bulletin of the American Astronomical Society, Vol. 33, p. 1043

**Scaramella, R.; Cascone, E.; Cortecchia, F.; et al.**, "GOHSS: a multi-echelle near IR spectrograph with OH subtraction", 2001, Conference Proceedings of *Scientific Dedication of the Telescopio Nazionale Galileo*, La Palma 3-5 November 2000, G. Setti, M.Rodonò, Eds.

**Schahmaneche, K.; Aldering, G.; Amanullah, R.; Antilogus, P.; Astier, P.; Balland, C.; Blanc, G.; Burns, M. S.; Conley, A.; Deustua, S.; et al.**, "Results from Recent High-redshift Type Ia Supernovae Searches", 2001, American Astronomical Society, 199th AAS Meeting, #16.12; Bulletin of the American Astronomical Society, Vol. 33, p. 1333

**Setti, G.**, "The Italian "Telescopio Nazionale Galileo", 2001, Conference Proceedings of *Scientific Dedication of the Telescopio Nazionale Galileo*, La Palma 3-5 November 2000, G. Setti, M. Rodonò, Eds.

**Viotti, R.; Altamore, A.; Rossi, C.**, "Exploring the nature of the symbiotic stars with high resolution spectroscopy", 2001, Mem. SAIt, 72, 753 - 756

**Vuerli, C.; Pasian, F.; Pucillo, M.; Smareglia, R.**, "The workstation software system", 2001, Conference Proceedings of *Scientific Dedication of the Telescopio Nazionale Galileo*, La Palma 3-5 November 2000, G. Setti, M. Rodonò, Eds.

# Publications **2000** on international journals with "referee"

**Duerbeck, H. W.; Liller, W.; Sterken, C.; Benetti, S.; Genderen, A. M.; Arts, J. M., Kurk, J.; Janson, M.; Voskes, T.; Brogt, E.; Arentoft, T.; van der Meer, A.; Dijkstra, R.**, "The rise and fall off V4334 Sgr (Sakurai's Object)", 2000, AJ, 119, 2360 (partly based on observations made with OIG@TNG)

**Klose, S.; Stecklum, B.; Masetti, N.; Pian, E.; Palazzi, E.; Henden, A. A.; Hartmann, D. H.; Fischer, O.; Gorosabel, J.; Sanchez-Fernandez, C.; et al.**, "The Very Red Afterglow of GRB 000418: Further Evidence for Dust Extinction in a Gamma-Ray Burst Host Galaxy", 2000, ApJ, 545, 271 (partly based on observations made with OIG@TNG)

**Maiolino, R.; Salvati, M.; Antonelli, L. A.; Comastri, A.; Fiore, F.; Ghinassi, F.; Gilli, R.; La Franca, F.; Mannucci, F.; Risaliti, G.; Thompson, D.; Vignali, C.**, "Optically dim counterparts of hard X-ray selected AGNs", 2000, A&A, 355, L47 (based on observations made with ARNICA@TNG)

**Masetti, N.; Bartolini, C.; Bernabei, S.; Guarnieri, A.; Palazzi, E.; Pian, E.; Piccioni, A.; Castro-Tirado, A. J.; Castro Ceron, J. M.; Verdes-Montenegro, L.; et al.**, "Unusually rapid variability of the GRB000301C optical afterglow", 2000, A&A, 359, 23 (partly based on observations made with OIG@TNG)

**Ragazzoni, R.; Baruffolo, A.; Marchetti, E.; Ghedina, A.; Farinato, J.; Niero, T.**, "Speckle interferometry measurements of the asteroids 10-Hygiea and 15-Eunomia", 2000, A&A, 354, 315 (based on observations made with the SPECKLE CAMERA@TNG)

**Ragazzoni, R.; Marchetti, E.; Valente, G.**, "Adaptive-optics corrections available for the whole sky", 2000, Nature, 403, 54 (based on data taken with the Rotator/Adapter acquisition probe optics of the TNG)

**Sabbadin, F.; Benetti, S.; Cappellaro, E.; Turatto, M.**, "The tetra-lobed planetary nebula NGC 1501", 2000, A&A, 361, 1112 (partly based on observations made with OIG@TNG)

**Sabbadin, F.; Cappellaro, E.; Benetti, S.; Turatto, M.; Zanin, C.**, "Tomography of the low excitation planetary nebula NGC 40", 2000, A&A, 355, 688 (partly based on observations made with OIG@TNG)

**Vignali, C.; Mignoli, M.; Comastri, A.; Maiolino, R.; Fiore, F.**, "Optical, near-

infrared and hard X-ray observations of SAXJ1353.9+1820: a red quasar", 2000, MNRAS, 314, L11 (partly based on observations made with OIG and ARNICA@TNG)

**Vuerli, C.; Pasian, F.; Pucillo, M.; Smareglia, R.**, "The High-Level Control and Data Handling System of the Galileo Telescope", 2000, Baltic Astronomy, v.9, p.523-526

# Publications 2000 on international non-refereed journals

**Altamore, A.; Baratta, G. B.; Cassatella, A.; Rossi, C.; Viotti, R.**, "Nature of symbiotic and related systems", 2000, *SARG at TNG: Prespectives for the Year 2000*, Proceedings of the workshop held in Padova (Italy), December 20-21, 1999, edited by R. G. Gratton and R. U. Claudi, p. 123

**Baffa, C.; et al.**, "The data acquisition system for NICS, the TNG infrared camera-spectrometer, software solutions", 2000, Proceedings of the International Meeting on Astronomical Technologies, 1999, *Telescopes, instruments and data processing for Astronomy in the year 2000*, S. Agata (Napoli), May 12-15, 1999, Mem. SAIt, Vol. 71, Supplement

**Benetti, S.; Cosentino, R.; Licandro, J.; Paulli, F.; Pedani, M.; Trancho, G.; Zacchei, A.; et al.**, "Supernova 2000ck in IC 4355", 2000, IAU Circular 7434, 1

**Benetti, S.; Dominguez, R. M.; Zacchei, A.; Molinari, E.; Giro, E.**, "Supernova 2000cu in ESO 525-G004", 2000, IAU Circular 7454, 2

**Benetti, S.; Zacchei, A.; Perez, I.; Filippenko, A. V.; Li, W. D.**, "Supernova 2000dg in MCG +1-1-29", 2000, IAU Circular 7484, 3

**Benetti, S.; Zacchei, A.; Perez, I.; Pedani, M.; Buzzoni, A.**, "Supernova 2000dj in NGC 735", 2000, IAU Circular 7491, 1

**Bonanno, G.**, "SARG CCD mosaic", 2000, *SARG at TNG: Prespectives for the Year 2000*, Proceedings of the workshop held in Padova (Italy), December 20-21, 1999, edited by R. G. Gratton and R. U. Claudi, p. 33

**Bonanno, G.; Bruno, P.; Cosentino, R.; Scuderi, S.; Bortoletto, F.; D'Alessandro, M.; Bonoli, C.; Fantinel, D.; Carbone, A.; Evola, G.**, "TNG new generation CCD controller", 2000, *Optical Detectors for Astronomy II: State-of-the-Art at the Turn of*

*the Millenium.* 4th ESO CCD Workshop, 1999, held in Garching, Germany. Edited by P. Amico and J. W. Beletic. Published by Kluwer Academic Publishers, P. O. Box 17, 3300 AA Dordrecht, ISBN 0-7923-6536-4, The Netherlands, p. 389

**Bonanno, A.; Ventura, R.**, "Detections of solar-like oscillations with SARG", 2000, *SARG at TNG: Prespectives for the Year 2000*, Proceedings of the workshop held in Padova (Italy), December 20-21, 1999, edited by R. G. Gratton and R. U. Claudi, p. 79

**Bonifacio, P.; Centurion, M.; Molaro, P.; Vladilo, G.**, "Using SARG to measure primordial deuterium", 2000, *SARG at TNG: Prespectives for the Year 2000*, Proceedings of the workshop held in Padova (Italy), December 20-21, 1999, edited by R. G. Gratton and R. U. Claudi, p. 97

**Bragaglia, A.; Carretta, E.; Gratton, R.; Tosi, M.**, "Old open clusters as tracers of galactic evolution", 2000, *SARG at TNG: Prespectives for the Year 2000*, Proceedings of the workshop held in Padova (Italy), December 20-21, 1999, edited by R. G. Gratton and R. U. Claudi, p. 103

**Bruno, P.**, "SARG movimentation and control system", 2000, *SARG at TNG: Prespectives for the Year 2000*, Proceedings of the workshop held in Padova (Italy), December 20-21, 1999, edited by R. G. Gratton and R. U. Claudi, p. 27

**Carretta, E.; Gratton; R. G.; Bragaglia, A.; Sneden, C.**, "Deep mixing in globular cluster stars", 2000, *SARG at TNG: Prespectives for the Year 2000*, Proceedings of the workshop held in Padova (Italy), December 20-21, 1999, edited by R. G. Gratton and R. U. Claudi, p. 99

**Catalano, S.**, "Detection of magnetic fields in young cluster late-type stars", 2000, *SARG at TNG: Prespectives for the Year 2000*, Proceedings of the workshop held in Padova (Italy), December 20-21, 1999, edited by R. G. Gratton and R. U. Claudi, p. 81

**Claudi, R. U.; Gratton, R. G.; Rebeschini, M.; Farisato, G.; Martorana, G.; Bonanno, G.; Bruno, P.; et al.**, "SARG: Documentation and performances", 2000, *SARG at TNG: Prespectives for the Year 2000*, Proceedings of the workshop held in Padova (Italy), December 20-21, 1999, edited by R. G. Gratton and R. U. Claudi, p. 49

**Claudi, R. U.; et al.**, "Search of planets and asteroseismology: perspectives using SARG at TNG", 2000, Proceedings of the International Meeting on Astronomical Technologies, 1999, *Telescopes, instruments and data processing for Astronomy in* *the year 2000*, S. Agata (Napoli), May 12-15, 1999, Mem. SAIt, Vol. 71, Supplement

**Clementini, C.; Cacciari, C.**, "The Baade – Wesselink method applied to cluster RR Lyrae", 2000, *SARG at TNG: Prespectives for the Year 2000*, Proceedings of the workshop held in Padova (Italy), December 20-21, 1999, edited by R. G. Gratton and R. U. Claudi, p. 75

**Comoretto, G.; et al.**, "The data acquisition system for NICS, the TNG infrared camera-spectrometer, hardware solutions", 2000, Proceedings of the International Meeting on Astronomical Technologies, 1999, *Telescopes, instruments and data processing for Astronomy in the year 2000*, S. Agata (Napoli), May 12-15, 1999, Mem. SAIt, Vol. 71, Supplement

**Conconi, P.; et al.**, "DOLoRes: the low resolution for Galileo", 2000, Proceedings of the International Meeting on Astronomical Technologies, 1999, *Telescopes, instruments and data processing for Astronomy in the year 2000*, S. Agata (Napoli), May 12-15, 1999, Mem. SAIt, Vol. 71, Supplement

**Cosentino, R.**, "The slit viewer", 2000, *SARG at TNG: Prespectives for the Year 2000*, Proceedings of the workshop held in Padova (Italy), December 20-21, 1999, edited by R. G. Gratton and R. U. Claudi, p. 45

**Cosentino, R.; Bonanno, G.; Bruno, P.; Calì, A.; Scuderi, S.; Timpanaro, M. C.**, "SARG control system", 2000, Proceedings of the International Meeting on Astronomical Technologies, 1999, *Telescopes, instruments and data processing for Astronomy in the year 2000*, S. Agata (Napoli), May 12-15, 1999, Mem. SAIt, Vol. 71, Supplement

**Cosentino, R.; Bonanno, G.; Bruno, P.; Scuderi, S.; Bonoli, C.; Bortoletto, F.; D'Alessandro, M.; Fantinel, D.**, "CCDs for the Instrumentation of the Telescopio Nazionale Galileo", 2000, *Further Developments in Scientific Optical Imaging*, proceedings of the International Conference on Scientific Optical Imaging held in Georgetown, Grand Cayman, on 2-5 December, 1998. Edited by M. Bonner Denton. Cambridge: Royal Society of Chemistry, p. 40

**Cosentino, R.; Bonanno, G.; Bruno, P.; Scuderi, S.; Bonoli, C.; Bortoletto, F.; D'Alessandro, M.; Fantinel, D.**, "CCDs for the instrumentation of the Telescopio Nazionale Galileo", 2000, Proceedings of the International Meeting on Astronomical Technologies, 1999, *Telescopes, instruments and data processing for Astronomy in the year 2000*, S. Agata (Napoli), May 12-15, 1999, Mem. SAIt, Vol. 71, Supplement

**Cremonese, G.**, "Sodium in the solar system", 2000, *SARG at TNG: Prespectives for*

*the Year 2000*, Proceedings of the workshop held in Padova (Italy), December 20-21, 1999, edited by R. G. Gratton and R. U. Claudi, p. 119

**Desidera, S.; Gratton, R. G.; Claudi, R. U.**, "Search for extrasolar planets using SARG", 2000, *SARG at TNG: Prespectives for the Year 2000*, Proceedings of the workshop held in Padova (Italy), December 20-21, 1999, edited by R. G. Gratton and R. U. Claudi, p. 73

**Desidera, S.; Gratton, R. G.; Claudi, R. U.; Rebeschini, M.; Farisato, G.; Calì, A.; Favero, G.**, "High precision radial velocities with iodine absorbing cells", 2000, *SARG at TNG: Prespectives for the Year 2000*, Proceedings of the workshop held in Padova (Italy), December 20-21, 1999, edited by R. G. Gratton and R. U. Claudi, p. 61

**Doressoundiram, A.; Barucci, M. A.; Romon, J.**, "Multi-color photometry of trans-Neptunian objects", 2000, American Astronomical Society, DPS Meeting #32, #21.10; Bulletin of the American Astronomical Society, Vol. 32, p. 1032

**Frasca, A.**, "Line asymmetries and surface convective motions in late-type stars", Detections of solar-like oscillations with SARG", 2000, *SARG at TNG: Prespectives for the Year 2000*, Proceedings of the workshop held in Padova (Italy), December 20-21, 1999, edited by R. G. Gratton and R. U. Claudi, p. 83

**Gardiol, D.; Bonoli, C.; Corcione, L.; Giro, E.; Zacchei, A.**, "Pointing and tracking software for the TNG telescope", 2000, Proc. SPIE Vol. 4009, p. 80-87, *Advanced Telescope and Instrumentation Control Software*, H. Lewis; Ed.

**Gratton, R. G.; Bonanno, G.; Bruno, P.; Cali, A.; Claudi, R. U.; Cosentino, R.; Desidera, S.; et al.**, "Tests of SARG: the high-resolution spectrograph for TNG", 2000, Proc. SPIE Vol. 4008, p. 244-255, *Optical and IR Telescope Instrumentation and Detectors*, M. Iye; A. F. Moorwood; Eds.

**Gratton, R. G.; Claudi, R. U.; Farisato, G.; Martorana, G.; Rebeschini, M.; et al.**, "SARG: the high resolution spectrograph of TNG", 2000, *SARG at TNG: Prespectives for the Year 2000*, Proceedings of the workshop held in Padova at the Padova Astronomical Observatory, December 20-21, 1999. Edited by R. G. Gratton and R. U. Claudi. Published by the Astronomical Observatory of Padova, Vicolo dell'Osservatorio 5, I-35122 Padova, Italy, p. 11

**Gratton, R. G.; Claudi, R.; Rebeschini, M.; Martorana, G.; Farisato, G.; Bonanno, G.; Bruno, P.; Cosentino, R.; Scuderi, S.**, "The high resolution spectrograph of TNG", 2000, Proceedings of the International Meeting on

Astronomical Technologies, 1999, *Telescopes, instruments and data processing for Astronomy in the year 2000*, S. Agata (Napoli), May 12-15, 1999, Mem. SAIt, Vol. 71, Supplement

**Held, E. V.; Ciattaglia, S. C.; Giro, E.; Zitelli, V.**, "Mask-Mode MOS at the TNG: a flexible approach to multi-object spectroscopy", 2000, Proceedings of the International Meeting on Astronomical Technologies, 1999, *Telescopes, instruments and data processing for Astronomy in the year 2000*, S. Agata (Napoli), May 12-15, 1999, Mem. SAIt, Vol. 71, Supplement

**Lanza, A. F.**, "Measuring stellar rotation and differential rotation in solar-type stars with SARG", 2000, *SARG at TNG: Prespectives for the Year 2000*, Proceedings of the workshop held in Padova (Italy), December 20-21, 1999, edited by R. G. Gratton and R. U. Claudi, p. 85

**Lanzafame, A.C.**, "Active late-type stars variability and models from high resolution spectra", 2000, *SARG at TNG: Prespectives for the Year 2000*, Proceedings of the workshop held in Padova (Italy), December 20-21, 1999, edited by R. G. Gratton and R. U. Claudi, p. 87

**Leone, F.**, "High resolution and stellar magnetic fields", 2000, *SARG at TNG: Prespectives for the Year 2000*, Proceedings of the workshop held in Padova (Italy), December 20-21, 1999, edited by R. G. Gratton and R. U. Claudi, p. 89

**Licandro, J.; Tessicini, G.; Perez, I.; Hidalgo, S.**, "Comet C/1999 S4 (LINEAR)", 2000, IAU Circular 7468, 2

**Lisi, F.; et al.**, "NICS, the IR imager / spectrograph of the TNG", 2000, Proceedings of the International Meeting on Astronomical Technologies, 1999, *Telescopes, instruments and data processing for Astronomy in the year 2000*, S. Agata (Napoli), May 12-15, 1999, Mem. SAIt, Vol. 71, Supplement

**Lorenzetti, D.; Cascone, E.; Cortecchia, F.; Ellis, R. S.; et al.**, "A progress report on GOHSS: a fiber-fed multiobject spectrograph for the Italian Galileo Telescope", 2000, Proceedings of the International Meeting on Astronomical Technologies, 1999, *Telescopes, instruments and data processing for Astronomy in the year 2000*, S. Agata (Napoli), May 12-15, 1999, Mem. SAIt, Vol. 71, Supplement

**Lorenzetti, D.; Cortecchia, F.; Vitali, F.; et al.**, "GOHSS (Galileo OH subtracted spectrograph): a progress report", 2000, Proc. SPIE Vol. 4008, p. 703-713, *Optical and IR Telescope Instrumentation and Detectors*, M. Iye; A. F. Moorwood; Eds.

**Mancini, D.; Schipani, P.**, "TNG motion control system", 2000, Proc. SPIE Vol. 4004, p. 277-280, *Telescope Structures, Enclosures, Controls, Assembly/Integration/Validation, and Commissioning*, T. A. Sebring; T. Andersen; Eds.

**Mancini, D.; Schipani, P.**, "Tracking performance of the TNG telescope", 2000, Proc. SPIE Vol. 4009, p. 355-365, *Advanced Telescope and Instrumentation Control Software*, H. Lewis; Ed.

**Mancini, D.; et al.**, "TNG: a progress report on the TNG drive and control system updates and on telescope performances", 2000, Proceedings of the International Meeting on Astronomical Technologies, 1999, *Telescopes, instruments and data processing for Astronomy in the year 2000*, S. Agata (Napoli), May 12-15, 1999, Mem. SAIt, Vol. 71, Supplement

**Mancini, D.; et al.**, "TNG: telescope control system organization", 2000, Proceedings of the International Meeting on Astronomical Technologies, 1999, *Telescopes, instruments and data processing for Astronomy in the year 2000*, S. Agata (Napoli), May 12-15, 1999, Mem. SAIt, Vol. 71, Supplement

**Mantegazza, L.**, "Asteroseismology of δ Scuti stars", 2000, *SARG at TNG: Prespectives for the Year 2000*, Proceedings of the workshop held in Padova (Italy), December 20-21, 1999, edited by R. G. Gratton and R. U. Claudi, p. 77

**Marilli, E.; Lanza, A. F.; Frasca, A.; Leto, G.**, "High resolution spectroscopy of τ Bootis: a star with a planet", 2000, *SARG at TNG: Prespectives for the Year 2000*, Proceedings of the workshop held in Padova (Italy), December 20-21, 1999, edited by R. G. Gratton and R. U. Claudi, p. 93

**Masetti, N.; Palazzi, E.; Pian, E.; Cosentino, R.; Ghinassi, F.; Magazzù, A.; Benetti, S.**, "GRB000812, TNG r-band observations", 2000, GRB Coordinates Network, Circular Service, 774, 1

**Micela, G.; Favata, F.; Sciortino, S.**, "The origin of Lithium in normal late K dwarfs", 2000, *SARG at TNG: Prespectives for the Year 2000*, Proceedings of the workshop held in Padova (Italy), December 20-21, 1999, edited by R. G. Gratton and R. U. Claudi, p. 125

**Micela, G.; Favata, F.; Sciortino, S.**, "Tracing the star formation history in Taurus-Auriga", 2000, *SARG at TNG: Prespectives for the Year 2000*, Proceedings of the workshop held in Padova (Italy), December 20-21, 1999, edited by R. G. Gratton and R. U. Claudi, p. 105

**Munari, U.**, "Internal kinematics, binaries and galactic motion of young stellar aggregates", 2000, *SARG at TNG: Prespectives for the Year 2000*, Proceedings of the workshop held in Padova (Italy), December 20-21, 1999, edited by R. G. Gratton and R. U. Claudi, p. 95

**Oliva, E.**, "Infrared instrumentation for large telescopes: an alternative approach", 2000, Mem SAIt, 71, 861

**Pagano, I.**, "Doppler imaging of late-type stars", 2000, *SARG at TNG: Prespectives for the Year 2000*, Proceedings of the workshop held in Padova (Italy), December 20-21, 1999, edited by R. G. Gratton and R. U. Claudi, p. 91

**Palazzi, E.; Masetti, N.; Magazzù, A.; Pedani, M.; Benetti, S.; Pian, E.**, "GRB000830, TNG b-band observations", 2000, GRB Coordinates Network, Circular Service, 786, 1

**Pallavicini, R.; Pasquini, L.; Randich, S.**, "Lithium in open clusters", 2000, *SARG at TNG: Prespectives for the Year 2000*, Proceedings of the workshop held in Padova (Italy), December 20-21, 1999, edited by R. G. Gratton and R. U. Claudi, p. 107

**Perinotto, M.**, "Planetary nebulae with SARG of TNG", 2000, *SARG at TNG: Prespectives for the Year 2000*, Proceedings of the workshop held in Padova (Italy), December 20-21, 1999, edited by R. G. Gratton and R. U. Claudi, p. 121

**Pernechele, C.; Bortoletto, F.; Conconi, P.; Gardiol, D.; Molinari, E.; Zerbi, F. M.**, "Preliminary design of a NIR prime focus corrector for the Galileo Telescope", 2000, Proc. SPIE Vol. 4008, p. 907-913, *Optical and IR Telescope Instrumentation and Detectors*, M. Iye; A. F. Moorwood; Eds.

**Pernechele, C.; Bortoletto, F.; Gardiol, D.; Ghedina, A.; Marchetti, E.**, "Ultimate test results on the active optics system of the Galileo Telescope", 2000, Proc. SPIE Vol. 4003, p. 116-121, *Optical Design, Materials, Fabrication, and Maintenance*, P. Dierickx; Ed.

**Pian, E.; Masetti, N.; Palazzi, E.; Frontera, F.; Ghinassi, F.; Licandro, J.; Gandolfi, G.**, "GRB000615: H-band observations", 2000, GRB Coordinates Network, Circular Service, 727, 1

**Piotto, G.; Recio Blanco, A.**, "Stellar rotation along the horizontal branches of globular clusters", 2000, *SARG at TNG: Prespectives for the Year 2000*, Proceedings of the workshop held in Padova (Italy), December 20-21, 1999, edited by R. G. Gratton and R. U. Claudi, p. 101

**Ragazzoni, R.; Baruffolo, A.; Farinato, J.; Ghedina, A.; Marchetti, E.; Esposito, S.**; et al., "Final commissioning phase of the AdOpt@TNG module", 2000, Proc. SPIE Vol. 4007, p. 57-62, *Adaptive Optical Systems Technology*, P. L. Wizinowich; Ed.

**Ragazzoni, R.; Baruffolo, A.; Marchetti, E.; Farinato, J.; Ghedina, A.; et al.**, "Mounting and testing AdOpt@TNG on the Canary sky", 2000, Proceedings of the International Meeting on Astronomical Technologies, 1999, *Telescopes, instruments and data processing for Astronomy in the year 2000*, S. Agata (Napoli), May 12-15, 1999, Mem. SAIt, Vol. 71, Supplement

**Ragazzoni, R.; Ghedina, A.; Baruffolo, A.; Marchetti, E.; Farinato, J.; Niero, T.; Crimi, G.; Ghigo, M.**, "Testing the pyramid wavefront sensor on the sky", 2000, Proc. SPIE Vol. 4007, p. 423-430, *Adaptive Optical Systems Technology*, P. L. Wizinowich; Ed.

**Rizzi, L.; Cappellaro, E.; Turatto, M.**, "High resolution spectroscopy of supernovae", 2000, *SARG at TNG: Prespectives for the Year 2000*, Proceedings of the workshop held in Padova (Italy), December 20-21, 1999, edited by R. G. Gratton and R. U. Claudi, p. 113

**Sabbadin, F.**, "Tomography of selected planetary nebulae", 2000, *SARG at TNG: Prespectives for the Year 2000*, Proceedings of the workshop held in Padova (Italy), December 20-21, 1999, edited by R. G. Gratton and R. U. Claudi, p. 127

**Scuderi, S.**, "SARG UIF", 2000, *SARG at TNG: Prespectives for the Year 2000*, Proceedings of the workshop held in Padova (Italy), December 20-21, 1999, edited by R. G. Gratton and R. U. Claudi, p. 43

**Scuderi, S.; Bonanno, G.; Bruno, P.; Calì, A.; Cosentino, R.**, "OIG and SARG CCD's characterization", 2000, Proceedings of the International Meeting on Astronomical Technologies, 1999, *Telescopes, instruments and data processing for Astronomy in the year 2000*, S. Agata (Napoli), May 12-15, 1999, Mem. SAIt, Vol. 71, Supplement

**Smareglia, R.; Pasian, F.; Vuerli, C.; Zacchei, A.**, "Operating the TNG Data

Handling and Archiving", 2000, *Astronomical Data Analysis Software and Systems IX*, ASP Conference Proceedings, Vol. 216, edited by N. Manset, C. Veillet, and D. Crabtree. Astronomical Society of the Pacific, ISBN 1-58381-047-1, p. 149

**Tagliaferri, G.; Cutispoto, G.; Pallavicini, R.; et al.**, "High resolution spectroscopy of X-ray serendipitous sources", 2000, *SARG at TNG: Prespectives for the Year 2000*, Proceedings of the workshop held in Padova (Italy), December 20-21, 1999, edited by R. G. Gratton and R. U. Claudi, p. 115

**Viotti, R.; Rossi, C.; Polcaro, V. F.; Norci, L.**, "Chemical evolution of hot massive stars", 2000, *SARG at TNG: Prespectives for the Year 2000*, Proceedings of the workshop held in Padova (Italy), December 20-21, 1999, edited by R. G. Gratton and R. U. Claudi, p. 117

**Vitali, F.; Cianci, E.; Lorenzetti, D.; Foglietti, V.; Notargiacomo, A.; Giovine, E.; Oliva, E.**, "Silicon grisms for high-resolution spectroscopy in the near infrared", 2000, Proc. SPIE Vol. 4008, p. 1383-1394, *Optical and IR Telescope Instrumentation and Detectors*, M. Iye; A. F. Moorwood; Eds.

**Zerbi, F. M.; Bortoletto, F.; Conconi, P.; Gardiol, D.; Molinari, E.; Pernechele, C.; Rizzoli, D.**, "Conceptual design for a NIR prime focus camera for the ESO 3.6", 2000, Proc. SPIE Vol. 4008, p. 822-829, *Optical and IR Telescope Instrumentation and Detectors*, M. Iye; A. F. Moorwood; Eds.

# Publications 1999 on international non-refereed journals

**Benetti, S.; Morossi, C.; Bortoletto, F.; Cosentino, R.; Gardiol, D.; Ghedina, A.; et al.**, "Supernova 1999cn in MCG +2-38-043", 1999, IAU Circular 7202, 1 (discovery of a supernova with OIG@TNG)

**Bonoli, C.; Vuerli, C.; Corcione, L.; Zacchei, A.; Gardiol, D.**, "Time base handling on the Galileo Telescope", 1999, *Instrumentation and Measurement Technology Conference*, 1999. IMTC/99. Proceedings of the 16th IEEE, vol. 2, p. 1228

**Bortoletto, F.**, "The TNG comissioning [Telescopio Nazionale Galileo]", 1999, *Instrumentation and Measurement Technology Conference*, IMTC/99. Proceedings of the 16th IEEE, Vol. 2, p. 627

**Bortoletto, F.; Bonoli, C.; Fantinel, D.; Gardiol, D.; Pernechele, C.**, "An active telescope secondary mirror control system", 1999, Review of Scientific Instruments (June 1999), Volume 70, Issue 6, pp. 2856-2860

**Gardiol, D.; Pernechele, C.**, "Online control of an active telescope secondary mirror", 1999, Proc. SPIE Vol. 3737, p. 601-607, *Design and Engineering of Optical Systems II*, F. Merkle; Ed.

**Ghedina, A.; Bortoletto, F.; Marchetti, E.; Gardiol, D.; Ragazzoni, R.; Pernechele, C.**, "The optics of Galileo Telescope: alignment and active optics preliminary results", 1999, *Instrumentation and Measurement Technology Conference*, IMTC/99. Proceedings of the 16th IEEE, vol. 2, p. 1223

**Mancini, D.; Schipani, P.**, "TNG rotator axes motion control", 1999, Proc. SPIE Vol. 3786, p. 369-375, *Optomechanical Engineering and Vibration Control*, E. A. Derby; C. G. Gordon; D. Vukobratovich; P. R. Yoder; C. Zweben; Eds.

**Mancini, D.; Schipani, P.**, "TNG main axes motion control progress report", 1999, Proc. SPIE Vol. 3786, p. 362-368, *Optomechanical Engineering and Vibration Control*, E. A. Derby; C. G. Gordon; D. Vukobratovich; P. R. Yoder; C. Zweben; Eds.

**Masetti, N.; Palazzi, E.; Pian, E.; Frontera, F.; Benetti, S.; Magazzù, A.; Castro-Tirado, A. J.; Jensen, B. L.**, "R-band observations of GRB991208 at TNG", 1999, GRB Coordinates Network, Circular Service, 462, 1

**Masetti, N.; Palazzi, E.; Pian, E.; Frontera, F.; Gardiol, D.; Benetti, S.; Zacchei,**

A.; et al., **"**X-ray/GRB 1SAX J0835.9+5118, TNG V and R observations", 1999, GRB Coordinates Network, Circular Service, 345, 1

**Molinari, E.; Conconi, P.; Pucillo, M.; Monai, S.**, "Galileo & DOLoRes", 1999, *Looking Deep in the Southern Sky*, Proceedings of the ESO/Australia Workshop held at Sydney, Australia, 10-12 December 1997. Edited by R. Morganti and W. J. Couch. Berlin: Springer-Verlag, p. 157.

**Pernechele, C.; Bortoletto, F.; Cavazza, A.; et al.**, "Optical alignment of the Galileo telescope: results and on-sky test before active optics final tuning", 1999, Proc. SPIE Vol. 3737, p. 594-600, *Design and Engineering of Optical Systems II*, F. Merkle; Ed.

**Ragazzoni, R.; Baruffolo, A.; Farinato, J.; Ghedina, A.; Marchetti, E.; Niero, T.**, "Toward Adopt@TNG First Light", 1999, *Astronomy with adaptive optics: present results and future programs*, Proceedings of an ESO/OSA topical meeting, held September 7-11, 1998, Sonthofen, Germany, Publisher: Garching, Germany: European Southern Observatory, ESO Conference and Workshop Proceedings, vol. 56, Edited by D. Bonaccini, p. 651

# Publications **1998** on international journals with "referee"

**Buffa, F.; Cocco, G. C.; Porceddu, I.; Serrau, M.**, "Dome seeing and temperature forecasting: a feasibility study for the Galileo Telescope", 1998, New Astronomy Reviews, Volume 42, Issue 6-8, p. 447-449

**Mancini, D.; Auricchio, A.; Brescia, M.; Ortolani, S.; Porceddu, I.; Sansone, M. P.; Zitelli, V.**, "The Galileo Telescope seeing monitor: technical overview and first results", 1998, New Astronomy Reviews, Volume 42, Issue 6-8, p. 425-429

# Publications **1998** on international non-refereed journals

**Baruffolo, A.; Ragazzoni, R.; Farinato, J.**, "AdOpt@TNG control system software", 1998, Proc. SPIE Vol. 3353, p. 1138-1145, *Adaptive Optical System Technologies*, D. Bonaccini; R. K. Tyson; Eds.

**Bortoletto, F.; Bonoli, C.; D'Alessandro, M.; Ragazzoni, R.; Conconi, P.; Mancini, D.; Pucillo, M.**, "Commissioning of the Italian National Telescope Galileo", 1998, Proc. SPIE Vol. 3352, p. 91-101, *Advanced Technology Optical/IR Telescopes VI*, L. M. Stepp; Ed.

**Ghedina, A.; Ragazzoni, R.; Baruffolo, A.; Farinato, J.**, "Low cost seeing monitor to measure the isokinetic patch on the edge of the moon", 1998, Proc. SPIE Vol. 3219, p. 73-82, *Optics in Atmospheric Propagation and Adaptive Systems II*, A. Kohnle; A. D. Devir; Eds.

**Gratton, R. G.; Cavazza, A.; Claudi, R. U.; Rebeschini, M.; Bonanno, G.; Bruno, P.; Cali, A.; Scuderi, S.; Cosentino, R.; Desidera, S.**, "High-resolution spectrograph of TNG: a status report", 1998, Proc. SPIE Vol. 3355, p. 769-776, *Optical Astronomical Instrumentation*, S. D'Odorico; Ed.

**Mancini, D.; Brescia, M.; Cascone, E.; Fiume, V.; Mancini, G.; Schipani, P.**, "Galileo Italian National Telescope (TNG) control system: adaptive preload control improvements", 1998, Proc. SPIE Vol. 3351, p. 135-138, *Telescope Control Systems III*, H. Lewis; Ed.

**Pasian, F.; Marcucci, P.; Pucillo, M.; Vuerli, C.; Malkov, O. Y.; Smirnov, O. M.; Monai, S.; Conconi, P.; Molinari, E.**, "Integrating the ZGSC and PPM at the Galileo Telescope for On-line Control of Instrumentation", 1998, *Astronomical Data Analysis Software and Systems VII*, ASP Conference Series, Vol. 145, 1998, R. Albrecht, R.N. Hook and H.A. Bushouse, eds., p. 433

**Pedichini, F.; Comari, M.; Cosentino, R.; Farisato, G.; de Baco, W.; Sandre, G.**, "High Precision Shutter for the T.N.G. CCD Camera", 1998, *Optical Detectors for Astronomy,* Proceedings of an ESO CCD workshop held in Garching, Germany, October 8-10, 1996. Edited by J. W. Beletic and P. Amico. Kluwer Academic Publishers, Boston, Mass. (Astrophysics and space science library; v. 228), p. 85

**Ragazzoni, R.; Baruffolo, A.; Farinato, J.; Ghedina, A.; Mallucci, S.; Marchetti, E.; Niero, T.**, "Final engineering test for AdOpt@TNG", 1998, Proc. SPIE Vol. 3353, p. 132-138, *Adaptive Optical System Technologies*, D. Bonaccini; R. K. Tyson; Eds.

**Riccardi, A.; Bindi, N.; Ragazzoni, R.; Esposito, S.; Stefanini, P.**, "Laboratory characterization of a Foucault-like wavefront sensor for adaptive optics", 1998, Proc. SPIE Vol. 3353, p. 941-951, *Adaptive Optical System Technologies*, D. Bonaccini; R. K. Tyson; Eds.

**Schipani, P.**, "TNG control system software architecture improvement", 1998, Proc. SPIE Vol. 3351, p. 165-171, *Telescope Control Systems III*, H. Lewis; Ed.

**Vuerli, C.; Bonoli, C.; Balestra, A.; Baruffolo, A.; Corcione, L.; et al.**, "Software integration at TNG and Active Optics: a practical example", 1998, Proc. SPIE Vol. 3351, p. 425-439, *Telescope Control Systems III*, H. Lewis; Ed.

# Publications **1997** on international journals with "referee"

**Barbieri, C.**, "The Galileo Italian National Telescope and its Instrumentation", 1997, Experimental Astronomy, v. 7, Issue 4, p. 257-279

**Pasian, F.; Smareglia, R.**, "The Data Flow, from Observations to the Archive: Simulating the Case of TNG", 1997, Experimental Astronomy, v. 7, Issue 4, p. 399-406

# Publications **1997** on international non-refereed journals

**Balestra, A.; Callegari, M.; Monai, S., et al.**, "Remote Control of the Galileo Telescope and the EU "REMOT" Project", 1997, *The Three Galileos: the Man, the Spacecraft, the Telescope*, Proceedings of the conference held in Padova, Italy on January 7-10, 1997 Publisher: Dordrecht Kluwer Academic Publishers, Astrophysics and space science library (ASSL) Series vol no 220, ISBN 0792348613, p. 371

**Balestra, A.; Pasian, F.; Pucillo, M.; Smareglia, R.; Vuerli, C.**, "Data Handling and Archiving at the Galileo Telescope", 1997, *The Three Galileos: the Man, the Spacecraft, the Telescope*, Proceedings of the conference held in Padova, Italy on January 7-10, 1997 Publisher: Dordrecht Kluwer Academic Publishers, Astrophysics and space science library (ASSL) Series vol no 220, ISBN 0792348613, p. 365

**Barbieri, C.**, "Galileo Italian National Telescope and its instrumentation", 1997, Proc. SPIE Vol. 2871, p. 244-255, *Optical Telescopes of Today and Tomorrow*, A. L. Ardeberg; Ed.

**Barbieri, C.**, "The Galileo Italian National Telescope and its Instrumentation", 1997, *The Three Galileos: the Man, the Spacecraft, the Telescope*, Proceedings of the conference held in Padova, Italy on January 7-10, 1997 Publisher: Dordrecht Kluwer Academic Publishers, Astrophysics and space science library (ASSL) Series vol no 220, ISBN 0792348613, p. 331

**Barbieri, C.**, "The Galileo Italian National Telescope", 1997, Mem SAIt, 68, 227

**Bortoletto, F.**, "The Galileo Telescope's Active Optics System", 1997, *The Three Galileos: the Man, the Spacecraft, the Telescope*, Proceedings of the conference held in Padova, Italy on January 7-10, 1997 Publisher: Dordrecht Kluwer Academic Publishers, Astrophysics and space science library (ASSL) Series vol no 220, ISBN 0792348613, p. 343

**Cascone, E.; Mancini, D.; Schipani, P.**, "Galileo Telescope model identification", 1997, Conference Paper, SPIE Proceedings, Vol. 3112, p. 343-350

**di Serego Alighieri, S.**, "Instrumentation and Observing Techniques from Galilei to the TNG", 1997, *The Three Galileos: the Man, the Spacecraft, the Telescope*, Proceedings of the conference held in Padova, Italy on January 7-10, 1997 Publisher: Dordrecht Kluwer Academic Publishers, Astrophysics and space science library (ASSL) Series vol no 220, ISBN 0792348613, p. 359

**Farinato, J.; Esposito, S.; Marchetti, E.; Ragazzoni, R.; Bruns, D. G.**, "Performance of a magnetic driven tip-tilt mirror", 1997, Proc. SPIE Vol. 2871, p. 962-968, *Optical Telescopes of Today and Tomorrow*, A. L. Ardeberg; Ed.

**Ghedina, A.; Ragazzoni, R.; Marchetti, E.**, "Optical design for the AdOpt@TNG module", 1997, Proc. SPIE Vol. 2871, p. 927-936, *Optical Telescopes of Today and Tomorrow*, A. L. Ardeberg; Ed.

**Gratton, R. G.; Bonanno, G.; Bhatia, R.; Cavazza, A.; Claudi, R. U.; Ferretti, F.**, "SARG: the high-resolution spectrograph of TNG", 1997, Proc. SPIE Vol. 2871, p. 1204-1215, *Optical Telescopes of Today and Tomorrow*, A. L. Ardeberg; Ed.

**Gratton, R. G., Cavazza, A.; Claudi, R. U.; et al.**, "SARG: The High Resolution Spectrograph of TNG", 1997, *The Three Galileos: the Man, the Spacecraft, the Telescope*, Proceedings of the conference held in Padova, Italy on January 7-10, 1997 Publisher: Dordrecht Kluwer Academic Publishers, Astrophysics and space science library (ASSL) Series vol no 220, ISBN 0792348613, p. 377

**Mancini, D.; Cascone, E.; Schipani, P.**, "Galileo high-resolution encoder system", 1997, Proc. SPIE Vol. 3112, p. 328-334, *Telescope Control Systems II*, H. Lewis; Ed.

**Mancini, D.; Cascone, E.; Schipani, P.**, "Italian National Galileo Telescope (TNG) system description and tracking performance in the workshop", 1997, Proc. SPIE Vol. 3086, p. 85-95, *Acquisition, Tracking, and Pointing XI*, M. K. Masten; L. A. Stockum; Eds.

**Marchetti, E.; Mallucci, S.; Ghedina, A.; Farinato, J.; Baruffolo, A.; Munari, U.; Ragazzoni, R.**, "A Real-Time Speckle Facility for the Telescopio Nazionale Galileo", 1997, *The Three Galileos: the Man, the Spacecraft, the Telescope*, Proceedings of the conference held in Padova, Italy on January 7-10, 1997 Publisher: Dordrecht Kluwer Academic Publishers, 1997 Astrophysics and space science library (ASSL) Series vol no 220, ISBN 0792348613, p. 383

**Marchetti, E.; Ragazzoni, R.; Farinato, J.; Ghedina, A.**, "Versatile wavefront simulator", 1997, Proc. SPIE Vol. 2871, p. 937-943, *Optical Telescopes of Today and Tomorrow*, A. L. Ardeberg; Ed.

**Masciadri, E.; Vernin, J.; Bougeault, P.**, "Seeing Prevision - A Possible Application to the TNG Telescope at La Palma", 1997, *The Three Galileos: the Man, the Spacecraft, the Telescope*, Proceedings of the conference held in Padova, Italy on January 7-10, 1997 Publisher: Dordrecht Kluwer Academic Publishers, Astrophysics and space science library (ASSL) Series vol n. 220, ISBN 0792348613, p. 389

**Molinari, E.; Conconi, P.; Pucillo, M.**, "LRS: a spectrograph for the TNG", 1997, Mem. SAIt, 68, 231

**Morossi, C.; Franchini, M.; Furlani, S.; Ragazzoni, R.; Marchetti, E.**, "Image selection by using an on-line fast shutter driven by tip-tilt signal", 1997, Proc. SPIE Vol. 2871, p. 897-904, *Optical Telescopes of Today and Tomorrow*, A. L. Ardeberg; Ed.

**Morossi, C.; Franchini, M.; Furlani, S.; Sedmak, G.**, "Online image selection for the Italian Telescopio Nazionale Galileo (TNG): numerical simulation and performance evaluation", 1997, Proc. SPIE Vol. 3126, *Adaptive Optics and Applications*, R. K. Tyson and R. Q. Fugate, Eds., p. 500

**Pernechele, C.; Bortoletto, F; Reif, K.**, "Position control for active secondary mirror of a two-mirror telescope", 1997, Proc. SPIE Vol. 3112, p. 172-180, *Telescope Control Systems II*, H. Lewis; Ed.

**Ragazzoni, R.**, "The Adaptive Optics Module for the Telescopio Nazionale Galileo", 1997, *The Three Galileos: the Man, the Spacecraft, the Telescope*, Proceedings of the conference held in Padova, Italy on January 7-10, 1997 Publisher: Dordrecht Kluwer Academic Publishers, Astrophysics and space science library (ASSL) Series vol no 220, ISBN 0792348613, p. 351

**Ragazzoni, R.**, "AdOpt@TNG. Adaptive optics at the Telescopio Nazionale Galileo. Yearly status report", 1997,  Edition December 1997 Publisher: Padova Osservatorio Astronomico, 1997 Physical description p.irr.. Electronic access at http//www.pd.astro.it/AdOpt/

**Ragazzoni, R.; Baruffolo, A.; Bortoletto, F.; D'Alessandro, M.; Farinato, J.; Ghedina, A.; Marchetti, E.**, "Adaptive optics module for TNG (AdOpt@TNG): a status report", 1997, Proc. SPIE Vol. 2871, p. 905-909, *Optical Telescopes of Today and Tomorrow*, A. L. Ardeberg; Ed.

**Ragazzoni, R.; Marchetti, E.; Gallieni, W. W.**, "Laser projection system for TNG", 1997, Proc. SPIE Vol. 2871, p. 944-947, *Optical Telescopes of Today and Tomorrow*, A. L. Ardeberg; Ed.

**Scaramella, R.; Cascone, E.; Cortecchia, F.; et al.**, "GOHSS: a Fibre-Fed Multiobject NIR Spectrograph for the Italian Galileo Telescope", 1997, *Extragalactic Astronomy in the Infrared*. Edited by G. A. Mamon, T. Xuan Thuan, and J. Tran Thanh Van. Paris: Editions Frontieres, p. 557

# Publications 1989-1996 on international journals with "referee"

**Balestra, A.; Marcucci, P.; Pasian, F.; Pucillo, M.; Smareglia, R.; Vuerli, C.**, "Using NIR Tools as Interfaces to Help and Archive Systems at the TNG Telescope", 1995, Vistas in Astronomy, vol. 39, Issue 1, pp. 79-87

**Oliva, E.; Gennari, S.**, "Achromatic lens systems for near infrared instruments", 1995, A&AS, 114, 179

# Publications 1989-1996 on international non-refereed journals

**Barbieri, C.**, "The Galileo project. A 3.5 m Italian telescope facility", 1989, Astrophysics and Space Science, vol. 160, p. 119

**Barbieri, C.**, "The project Telescopio Nazionale Galileo", 1993, Mem. SAIt, 64, 657

**Barbieri, C.; Baruffolo, A.; Bhatia, R.; Bonoli, C.; Bortoletto, F.; Canton, G.; Ciani, A.; Conconi, P.; D'Alessandro, M.; Fantinel, D.; et al.**, "The Galileo Italian National Telescope", 1992, ESO Conference on *Progress in Telescope and Instrumentation Technologies*, ESO, Garching, 27-30 April 1992, Garching: European Southern Observatory (ESO), edited by M.-H. Ulrich, p. 137

**Barbieri, C.; Bhatia, R. K.; Bonoli, C.; Bortoletto, F.; Ciani, A.; Conconi, P.; et al.**, "Status of the Galileo National Telescope", 1994, Proc. SPIE Vol. 2199, p. 10-21, *Advanced Technology Optical Telescopes V*, L. M. Stepp; Ed.

**Barbieri, C.; Cristiani, S.; Zambon, M.**, "The Galileo telescope (Il telescopio Galileo)", 1990, Mem. SAIt, 61, 217

**Baruffolo, A.; Bortoletto, F.; Fantinel, D.; Bonoli, C.**, "Data Acquisition and Control Informatics for the Galileo Telescope", 1992, ESO Conference on *Progress in Telescope and Instrumentation Technologies*, ESO, Garching, 27-30 April 1992, Garching: European Southern Observatory (ESO), edited by M.-H. Ulrich, p. 349

**Bhatia, R. K.; Ciani, A.; Rafanelli, P.**, "Shack-Hartmann analysis of the 3.58 m primary mirror of the Galileo telescope", 1994, Proc. SPIE Vol. 1994, p. 68-78, *Advanced Optical Manufacturing and Testing IV*, V. J. Doherty; Ed.

**Bonaccini, D.; Esposito, S.; Brusa, G.**, "Adaptive optics with liquid crystal phase screens", 1994, Proc. SPIE Vol. 2201, p. 1155-1158, *Adaptive Optics in Astronomy*, M. A. Ealey; F. Merkle; Eds.

**Bonanno, G.; Bruno, P.; Cosentino, R.; et al.**, "Detector Controllers for the Galileo Telescope: A Progress Report", 1995, IAU Symposium No. 167, *New Developments in Array Technology and Applications* (1995), A. G. D. Philip, K. A. Janes and A. R. Upgren, eds., Kluwer Academic Publishers, Dordrecht. Held in the Hague, the Netherlands, August 23-27, 1994., p. 319

**Bonoli, C.; Mancini, D.; Bortoletto, F.; Corcione, L.; D'Alessandro, M.; Schipani, P.; Stefani, R.**, "TNG control system: computer architecture, interfacing, and synchronization", 1995, Proc. SPIE Vol. 2479, p. 160-168, *Telescope Control Systems*, P. T. Wallace; Ed.

**Bortoletto, F.; Baruffolo, A.; Bonoli, C.; et al.**, "Primary Mirror Control System for the Galileo Telescope", 1991, *Active and Adaptive Optical Systems*, SPIE 1542, p. 225

**Bortoletto, F.; Bonoli, C.; D'Alessandro, M.; Fantinel, D.; Farisato, G.; Bonanno, G.; Bruno, P.; Cosentino, R.; Bregoli, G.; Comari, M.**, "CCD cameras for the Italian national telescope Galileo", 1996, Proc. SPIE Vol. 2654, p. 248-258, *Solid State Sensor Arrays and CCD Cameras*, C. N. Anagnostopoulos; M. M. Blouke; M. P. Lesser; Eds.

**Bortoletto, F.; Fantinel, D.; Gallieni, W.; Giudici, G.; Ragazzoni, R.; Tommelleri, R.; Vanini, P.**, "Active Optics Control System for the Galileo Telescope. A Status Report", 1992, ESO Conference on *Progress in Telescope and Instrumentation Technologies*, ESO, Garching, 27-30 April 1992, Garching: European Southern Observatory (ESO), edited by M.-H. Ulrich, p. 323

**Bortoletto, F.; Fantinel, D.; Ragazzoni, R.; Bonoli, C.; D'Alessandro, M.; Balestra, A.; Marcucci, P.; Pucillo, M.; Vuerli, C.**, "Active optics handling inside Galileo Telescope", 1994, Proc. SPIE Vol. 2199, p. 212-222, *Advanced Technology Optical Telescopes V*, L. M. Stepp; Ed.

**Cavazza, A.; Bathia, R.; Gratton, R.**, "The Galileo Echelle Spectrograph and a Preliminary Design of the Red Camera", 1992, ESO Conference on *Progress in Telescope and Instrumentation Technologies*, ESO, Garching, 27-30 April 1992, Garching: European Southern Observatory (ESO), 1992, edited by M.-H. Ulrich, p. 741

**di Serego Alighieri, S.**, "Telescopio Nazionale Galileo", 1995,
ESO Conference and Workshop Proceedings, Proceedings of an ESO/ST-ECF
workshop on *Calibrating and Understanding HST and ESO instruments*, 25-28 April
1995, Garching, Germany, Garching near Munich: European Southern Observatory,
edited by P. Benvenuti, p. 221

**di Serego Alighieri, S.**, "Observing Modes for the Italian Galileo Telescope", 1996,
*New observing modes for the next century*. Astronomical Society of the Pacific
Conference Series, Volume 87, Proceedings of a workshop held in Hilo, Hawaii, 6-8
July 1995, San Francisco: Astronomical Society of the Pacific (ASP), |c1996, edited
by T. Boroson, J. Davies, and I. Robson, p. 200

**Gennari, S.; Vanzi, L.; Lisi, F.**, "Optical design of the infrared camera for the
Galileo Telescope (TNG)", 1995, Proc. SPIE Vol. 2475, p. 221-227, *Infrared
Detectors and Instrumentation for Astronomy*, A. M. Fowler; Ed.

**Gratton, R. G.; Bhatia, R.; Bonanno, G; et al.**, "A high resolution spectrograph for
the Galileo National Telescope", 1993, Mem. SAIt, 64, 672

**Gratton, R. G.**, "Scientific Programs for the High Resolution Spectrograph of the
Galileo National Telescope", 1994, Mem. SAIt, 65, 791

**Gratton, R. G.; Bhatia, R. K.; Cavazza, A.**, "High-resolution spectrograph for the
Galileo National Telescope", 1994, Proc. SPIE Vol. 2198, p. 309-316,
*Instrumentation in Astronomy VIII*, D. L. Crawford; E. R. Craine; Eds.

**Knohl, E.-D.; Schillke, F.; Schmidt, M.**, "Another Milestone in Modern Astronomy:
The 3.6m Primary Mirror of the Telescopio Nazionale Galileo (TNG)", 1992,
Proceedings of ESO Conference on *Progress in Telescope and Instrumentation
Technologies*, ESO, Garching, 27-30 April 1992, Garching: European Southern
Observatory (ESO), 1992, edited by M.-H. Ulrich, p. 129

**Lisi, F.**, "The infrared camera/spectrometer for the Galileo Telescope", 1993, Mem.
SAIt, 64, 669

**Mancini, D.**, "Italian National Galileo Telescope (TNG) control system: hardware,
software, and methods adopted to improve the performance of the fully digital drive
system", 1994, Proc. SPIE Vol. 2479, p. 245-252, *Telescope Control Systems*, P. T.
Wallace; Ed.

**Mancini, D.**, "Galileo Italian National Telescope (TNG) drive system: components, methods, and performances", 1994, Proc. SPIE Vol. 2199, p. 352-363, *Advanced Technology Optical Telescopes V*, L. M. Stepp; Ed.

**Pasian, F.**, "Archiving TNG Data", 1996, *Astronomical Data Analysis Software and Systems V*, ASP Conference Series, Vol. 101, G. H. Jacoby and J. Barnes, eds., p. 479.

**Pasian, F.; Pucillo, M.**, "The calibration archive and quick-reduction at the TNG telescope", 1995, ESO Conference and Workshop Proceedings, Proceedings of an ESO/ST-ECF workshop on *Calibrating and Understanding HST and ESO instruments*, 25-28 April 1995, Garching, Germany, Garching near Munich: European Southern Observatory, edited by P. Benvenuti, p. 237

**Ragazzoni, R.; Bonaccini, D.**, "The adaptive optics system for the Telescopio Nazionale Galileo", 1996, *Adaptive Optics*. ESO Conference and Workshop Proceedings, Proceedings of a topical meeting, held October 2-6, 1995, Garching, Germany, Garching near Munich: European Southern Observatory, edited by M. Cullum, p. 17

# List of TNG technical reports 1990 – 2000

1. **Barbieri, C.**, "The Galileo National Telescope", October 1990

2. **Barbieri, C.; Bortoletto, F.; D'Alessandro, M.; Ragazzoni, R.; Scotton, V.**, "2D photon counting at Padova-Asiago Observatory", September 1990

3. **Bhatia, R.K.; Gratton, R.; Iovino, A.**, "The Fourier Transform Spectrometer vs. Echelle debate revisited", October 1990

4. **Ragazzoni, R.**, "A preliminary design for auxiliary optics for HST-FOC", December 1990

5. **Bhatia, R.K.; Gratton, R.; Iovino, A.**, "Design considerations for an echelle spectrograph for the 3.5m Italian telescope Galileo", June 1991

6. **Baruffolo, A.; Bonoli, C.; Ciani, A.**, "The TNG command architecture", June 1991

7. **Bortoletto, F.; Baruffolo, A.; Bonoli, C.; D'Alessandro, M.; Fantinel, D.; Giudici, G.; Ragazzoni, R.; Salvadori, L.; Vanini, P.**, "Primary mirror control system for the GALILEO telescope", September 1991

8. **Ragazzoni, R.; Bortoletto, F.**, "Moving M2 mirror without pointing offset", September 1991

9. **Balestra, A.; Marcucci, P.; Pasian, F.; Pucillo, M.; Smareglia, R.; Vuerli, C.**, "Galileo Project Workstation Software System", December 1992

10. **Ragazzoni, R.**, "IDL-SH, A package of Shack-Hartmann data reduction under IDL environment", February 1992

11. **Bathia, R.; Ciani, A.**, "Active control of spectrum drifts in spectrographs", February 1992

12. **Gratton, R.; Bhatia, R.**, "Theory of the post-dispersed Fourier transform spectrographs", March 1992

13. **Barbieri, C.; Baruffolo, A.; Bhatia, R.; Bonoli, C.; Bortoletto, F.; Canton, G.; Ciani, A.; Conconi, P.; et al.**, "The Status of the Galileo National Telescope", June 1992

14. **Bonoli, C.; Fantinel, D.; Baruffolo, A.; Bortoletto, F.**, "A distributed VME Telescope Control System for Remote Operations", June 1992

15. **Marcucci, P.; Pucillo, M.**, "The Galileo Table Editor", October 1992

16. **Marcucci, P.**, "The Galileo Help System, Version 1.0", December 1992

17. **Balestra, A.**, "The Galileo Message Exchange System", December 1992

18. **Ragazzoni, R.**, "Optimization of auxiliary optics in Active Optics Telescopes", December 1992

19. **Pasian, F.; Pucillo, M.**, "TNG Data Handling and Archiving: an Overview", December 1992

20. **Bhatia, R.; Ciani, A.**, "Optical Quality of the Primary Mirror of the Galileo Telescope from the Shack--Hartmann Test", March 1993

21. **Mancini, D.**, "The TNG Main Axes Drive System", August 1993

22. **Mancini, D.**, "The GALILEO Telescope Power System", August 1993

23. **Bortoletto, F.; Fantinel, D.; Giudici, G.; Ragazzoni, R.**, "Testing Active Optics for the National Telescope GALILEO", August 1993

24. **Bhatia, R.; Ciani, A.; Rafanelli, P.**, "Results from the Shack--Hartmann analysis of the 3.5m primary mirror of the Galileo telescope", September 1993

25. **Ciani, A.; Bhatia, R.**, "Orthogonality of the Zernike Polynomials: an experimental investigation", September 1993

26. **Gratton, R.; Ciani, A.; Conconi, P.**, "Atmospheric Dispertion Corrector (ADC) for the TNG", September 1993

27. **Bevini, F.**, "Mechanical Backlash in the TNG Azimuth Feed Drives", September 1993

28. **Mancini, D.; Mancini, G.; Ortolani, S.; Zitelli, V.**, "The Seeing Monitor Tower", September 1993

29. **Bevini, F.**, "Hydrostatic Bearing System for the TNG Azimuth Axis", November 1993

30. **Ragazzoni, R.; Marchetti, E.**, "Raleigh vs. Sodium Beacon for Large FoV Partial Correction", February 1994

31. **Ragazzoni, R.; Marchetti, E.**, "A Liquid Adaptive Mirror", February 1994

32. **Ragazzoni, R.**, "Some useful relationships among optical parameters in the TNG", March 1994

33. **Bortoletto, F.; Fantinel, D.; Ragazzoni, R.; Bonoli, C.; D'Alessandro, M.; Balestra, A.; Marcucci, P.; Pucillo, M.; Vuerli, C.**, "Active optics handling inside Galileo Telescope", April 1994

34. **Bhatia, R.**, "Active spectrographs: a new way to improve their quality", April 1994

35. **Bhatia, R.; Ciani, A.**, "Defining mirror quality: a global approach", April 1994

36. **Barbieri, C.; Bhatia, R.; Bonoli, C.; Bortoletto, F.; Canton, G.; Ciani, A.; Conconi, P.; D'Alessandro, M.; Fantinel, D.; Mancini, D.; et al.**, "The Status of the Galileo National Telescope", May 1994

37. **Esposito, S.; Brusa, G.; Bonaccini, D.; Biliotti, V.**, "Laboratory results on tip/tilt compensation for astronomy", May 1994

38. **Bonaccini, D.**, "Adaptive Optics System Error Budgets and Parametric Analysis", May 1994

39. **Ragazzoni, R.**, "Optics for the TNG Rotator/Adapter", October 1994

40. **Bhatia, R.; Ciani, A.**, "Active Spectrographs: using the Shack-Hartmann principle to improve their quality", October 1994

41. **di Serego Alighieri, S.; Bonaccini, D.; Oliva, E.; Piotto, G.; Ragazzoni, R.; Richichi, A.**, "Adaptive Optics for the TNG", January 1995

42. **Ragazzoni, R.; Claudi, R.U.**, "An unusual aberration of very large Liquid Mirror Telescopes", March 1995

43. **Bhatia, R.**, "Telescope alignment: is the zero-coma condition sufficient?", May 1995

44. **Bonoli, C.; Mancini, D.; Bortoletto, F.; Corcione, L.; D'Alessandro, M.; Schipani, P.; Stefani, R.**, "TNG Control System: computer architecture, interfacing and synchronization", May 1995

45. **Marchetti, E.**, "Design of a Prime Focus Corrector for a 1.5m Class Telescope", May 1995

46. **Ragazzoni, R.**, "Pupil Plane Wavefront Sensing with an Oscillating Prism", August 1995

47. **Ragazzoni, R.; Esposito, S.; Marchetti, E.**, "Auxiliary Telescopes for the Absolute Tip-Tilt Determination of a Laser Guide Star", August 1995

48. **Cavazza, A.; Gratton, R.G.; Bhatia, R.K.**, "Optical Design of the TNG High Resolution Spectrograph (SARG): the R4 Option", September 1995

49. **Pasian, F.; Pucillo, M.**, "The Calibration Archive and Quick-Reduction at the TNG Telescope", September 1995

50. **Mancini, D.; Schipani, P.**, "The TNG Control System: the Encoder Subsystem", September 1995

51. **Pasian, F.**, "Archives at the TNG Telescope: Architectural Design Document",

September 1995

52. **Ragazzoni, R.**, "Absolute Tip-Tilt Determination with Laser Beacons", October 1995

53. **Barbieri, C.**, "The GALILEO Italian National Telescope and its Instrumentation", November 1995

54. **Fini, L.; Ranfagni, P.**, "The TNG Tip-Tilt Servo-Loop", December 1995

55. **Pasian, F.**, "Data Reduction Software for TNG Instruments: Definitions", January 1996

56. **Ragazzoni, R.**, "Propagation delay of a laser beacon as a tool to retrieve absolute tilt measurements", April 1996

57. **Esposito, S.; Riccardi, A.; Ragazzoni, R.**, "Focus anisoplanatism effects on tip-tilt compensation for Adaptive Optics using sodium laser beacons as tracking reference", July 1996

58. **Ragazzoni, R.; Marchetti, E.; Brusa, G.**, "The effective layer height for star wandering and the accuracy of tilt sensing in multicolour laser stars", July 1996

59. **Mancini, D.; Cascone, E.; Schipani, P.**, "Control system description and tracking performance of the TNG in the workshop", September 1996

60. **Marchetti, E.; Ragazzoni, R.**, "Sky coverage with the auxiliary telescopes LGS tilt recovery technique", January 1997

61. **Ragazzoni, R.**, "Robust tilt determination from Laser Guide Stars using a combination of different techniques", January 1997

62. **Ghedina, A.; Ragazzoni, R.**, "Optimum configurations for two off-axis parabolae optical relay", February 1997

63. **Vega, J.C.; Corsini, E.M.; Pizzella, A.; Bertola, F.**, "Figure-of-eight velocity curves: UGC 10205", February 1997

64. **Caporali, A.; Barbieri, C.**, "The Astronomic and Geodetic Coordinates of the Telescopio Nazionale Galileo, Canary Islands", April 1997

65. **Caporali, A.**, "Alignment of the TNG Dome relative to Trigonometric Verteces", April 1997

66. **Vitali, F.; Lorenzetti, D.; Ferruzzi, D.; Oliva, E.**, "The Infrared Grisms for the Galileo Telescope", July 1997

67. **Buffa, F.; Porceddu, I.**, "Temperature forcast and dome seeing minimization. I. A case study using a neural network model", October 1997

68. **Ragazzoni, R.**, "Laser Guide Star Advanced Concepts: Tilt Problem, Perspective Approach and Beyond", November 1997

69. **Ghedina, A.; Ragazzoni, R.; Baruffolo, A.**, "Isokinetic patch measurements on the edge of the Moon", December 1997

70. **Morossi, C.; Franchini, M., Furlani, S.; Sedmak, G.**, "On-line image selection for the Italian Telescopio Nazionale Galileo (TNG): numerical simulation and performance evaluation", December 1997

71. **Esposito, S.; Marchetti, E.; Ragazzoni, R.; Baruffolo, A.; Farinato, J.; Fini, L.; Ghedina, A.; Ranfagni, P.; Riccardi, A.**, "Laboratory characterization of an APD-based tip-tilt corrector", February 1998

72. **Gratton, R.; Cavazza, A.; Claudi, R.U.; Rebeschini, M.; Bonanno, G.; Bruno, P.; Cali, A.; Scuderi, S.; Cosentino, R.; Desidera, S.**, "The High Resolution Spectrograph of TNG: a Status Report", March 1998

73. **The TNG Commissioning Group**, "TNG Software Commissioning: Issue 1", March 1998

74. **Pernechele, C.**, "The TNG Active Optics User Interface", April 1998

75. **Gardiol, D.; Zacchei, A.**, "The TNG Telescope Control User Interface", April 1998

76. **Giro, E.; Pernechele, C.**, "The TNG Rotator Adapter User Interface", April 1998

77. **The TNG Commissioning Group**, "TNG Optics Commissioning: Issue 1", April 1998

78. **Pucillo, M.; Conconi, P.; Molinari, E.; Monai, S.**, "Low Resolution Spectrograph for the TNG: Functional Specifications for the Control Software", June 1998

79. **Comari, M.; Corte, C.**, "Test and Characterization of the Temperature Transducers for the CCD Cameras of the Italian National Telescope Galileo", June 1998

80. **Pernechele, C.**, "The TNG Wavefront Sensor User Interface: Software and simulations", July 1998

81. **Vitali, F.; Lorenzetti, D.**, "The K Band Grism for the Spectroscopic Mode of NICS@TNG", September 1999

82. **Mancini, D.; Schipani, P.**, "TNG control and drive system command handling", January 2000

# List of TNG newsletters 1992 – 1998

1. ***January 1992*** issue
   Contents: foreword - CRA and Galileo - a brief history of the italian national telescope - from blank to mirrors - TNG scientific instruments - an interview with F. Roddier - structure of the project office - our consultants - news from italian observatories - list of pubblications

2. ***August 1992*** issue
   Contents: the status of the project - a welcome from the canaries - TNG optics - ANTARES - adaptive optics - interview with R.L.Davies - the site and DIMM - the TNG building - homage to Galileo - spectrograph calibration

3. ***December 1992*** issue
   Contents: mechanical acceptance tests for the cell - M1 cell system - optical tests - the DIMM tower - instrumentation overview - new determination of Asiago coordinates

4. ***April 1993*** issue
   Contents: M1 presentation at Zeiss - the VME control network - our interviews - our consultants

5. ***September 1993*** issue
   Contents: TNG control system - CCD detectors - more on ANTARES - news in astronomy - a talk at the italian embassy - Galileo Galilei's telescopes

6. ***December 1993*** issue
   Contents: status report - site activities - transputers - ANTARES - CCD's update - CLUE experiment - our consultants - our interviews – news

7. ***April 1994*** issue
   Contents: latest developments - wiring harness - mechanical alignment - our interviews - news on Pluto

8. ***September 1994*** issue
   Contents: progress report - the TNG observatory - a measure of K - ANTARES update - TNG metereology I - GPS for the TNG - our interviews - our consultants

9. ***January 1995*** issue
   Contents: ceremony in Milano - TNG drive system - TNG metereology II - M2 revisited - multi-media - aligning the base plate

10. ***May 1995*** issue
    Contents: progress report - contract awarded for the rotating building - the TNG rotator adapters - mounting of the az-alt encoders - control system of the rotating building - our consultants and contractors - our interviews - what is LBT?

11. ***September 1995*** issue
    Contents: status report - adaptive optics - TNG building rotating system - TNG

UTM and astronomical positions - telescope structure: elastic qualification measurements - notice to the readers

12. *January 1996* issue
Contents: status report – installation and alignment of the rotation system – the TNG differential image motion monitor (DIMM) – light pollution meeting in Asiago – L. Rosino 80[th] birthday – our interviews

13. *May 1996* issue
Contents: status report - rotator adapter optics - drive, control and encoder system tests - the TNG low resolution spectrograph - our interview: M.Mendillo - lunar observations from the TNG site - ESO-TNG meeting - Galileian lectures - TNG to be dedicated

14. *September 1996* issue
Contents: june 29, 1996: inauguration ceremony - status report - 1st national competition for italian schools - geodetic and astronomical coordinates

15. *February 1997* issue
Contents: status report - the italian consortium for astronomy and astrophysics - the three Galileos conference - exhibition: voyage in the cosmos - rotator adapters overview - news: DOLoRes - our interviews: M.Marov - our consultants

16. *June 1997* issue
Contents: status report - astronomy and metereology III - neural meteorological forecasting - our interviews: P.L.Bernacca - news

17. *October 1997* issue
Contents: status report - the local area network at the TNG - aligning the TNG: definition of the main axes - CCD working group activity - news

18. *January 1998* issue
Contents: status report - CNAA headquarters inaugurated - our interviews: M.Rodono' - the "Centro Galileo Galilei" - adaptive optics update - our consultants – news

19. *May 1998* issue
Contents: status report - the TNG control user interface - active optics and rotator adapter user interfaces - DIMM news - the TNG weather station - TAC document - news

# List of  TNG technical documents of internal use

- **Tessicini, G.; Garcia de Gurtubai, A.; Mainella, G.; Pinilla Alonso, N.**, "Manual del telescopio", 2006, in progress

- **Ghedina, A.**, "Activity on M2: January 9$^{th}$ – 22$^{nd}$, 2006", January 2006

- **Tessicini, G.; et al.**, "Manuale delle emergenze", last update December 2005

- **Oliva, E.; Carmona, C.; Ghedina, A.; Gonzalez, C.; Gonzalez, M.; Huertas, M.; Lodi, M.; Merges, F.; Ventura, H.**, "πODer, the motor control system for the derotators optics", last update December 2005

- **Tessicini, G.; Ghedina, A.; Pedani, M.**, "Manuale delle emergenze meteo", last update November 2005

- **Oliva, E.; Scuderi, S.; Lodi, M.**, "Sensitivity of the autoguide cameras and selection of stars for autoguiding", November 2005

- **Huertas, M.**, "Poder graphical user interface", 2005

- **Huertas, M.**, "Telescope Control Panel: User Guide", 2005

- **Gonzalez, M.**, "Manual para el cambio de rabbit multiserial", 2005

- **Scuderi, S.**, "TNG Autoguide: GUI Manual", October 2005

- **Scuderi, S.**, "TNG Autoguide: Quick Reference Guide", October 2005

- **Oliva, E.; Carmona, C.; Gonzalez, M.; Schipani, P.**, "Status of the AZ and EL drive systems", April 2005

- **Ghinassi, F.**, "Manuale dell'utente NICS", last update March 2005

- **Oliva, E.; Scuderi, S.; Lodi, M.**, "Requirements for the new autoguide interface", January 2005 (updated July 2005)

- **Pedani, M.**, "Manuale d'uso della Workstation Maintenance di LRS", December 2004

- **Ghedina, A.**, "Manuale per lo smontaggio di M1", November 2004

- **Ghedina, A.**, "Manuale per il lavaggio di M1 ed M3", November 2004

- **AdOpt@TNG team**, "AdOpt@TNG user manual", October 2004

- **Pedani, M.**, "Manuale per il montaggio delle maschere MOS di LRS", October 2004

- **Oliva, E.; Lorenzi, V.,** "User guide for polarimetric observations with NICS", August 2004

- **Oliva, E.; Schipani, P.; Carmona, C.; Gonzales, M.,** "Status report of the azimuth, elevation and derotator encoders", last update April 2004

- **Scuderi, S.,** "SARG user interface manual", December 2003 (document SARG – D035)

- **Carmona, C.,** "Procedimiento y funcionamento de los compresores para acondicionamiento basados en R22 incluida carga de gas y tara de la valvula de expansion", March 2003

- **Gavryusev, V.,** "Xnics user guide", last update January 2003 (Technical Report 1/2003, IRA CNR, sez. Firenze)

- **Mannucci, F.,** "SNAP: Speedy Near-infrared data Automatic Pipeline", November 2002 (vers. 1.1)

- **Pedani, M.,** "The pointing of the Telescopio Nazionale Galileo", September 2002

- **Cosentino, R; Bruno, P.,** "TNG: The SARG derotator interface", November 2001

- **Oliva, E.,** "Derotator-B bearing failure and fixing in July 2001", October 2001 (updated on March 2003)

- **Di Fabrizio, L.,** "Analysis of the TRAM's movements and optical aberrations of DOLoRes as a function of the Nasmyth B derotator angle", May 2001

- **Ghinassi, F.; Cosentino, R.,** "TNG: Manuale di vuoto e criogenia", Febbraio 2001

- **Bonoli, C.,** "TNG: pointing and tracking manual", July 2000

- **Bortoletto, F.; et al.,** "The OIG user manual", June 2000

- **Bonoli, C.,** "TNG: the telescope user manual", March 2000